\documentclass{article}
\usepackage{jcappub}
\usepackage{epsfig}
\usepackage{graphicx}
\usepackage[latin1]{inputenc}
\usepackage{amsmath}
\usepackage{url}
\def\simlt{\ \raise -2.truept\hbox{\rlap{\hbox{$\sim$}}\raise5.truept   %
\hbox{$<$}\ }}
\def\simgt{\ \raise -2.truept\hbox{\rlap{\hbox{$\sim$}}\raise5.truept   %
\hbox{$>$}\ }}                                                          %

\def\be{\begin{equation}}
\def\ee{\end{equation}}
\def\newline{\hfil\break}

\def\la{\mathrel{\hbox{\rlap{\hbox{\lower4pt\hbox{$\sim$}}}\hbox{$<$}}}}
\def\ga{\mathrel{\hbox{\rlap{\hbox{\lower4pt\hbox{$\sim$}}}\hbox{$>$}}}}

\title{Polarization of the Sunyaev-Zel'dovich effect: relativistic imprint of thermal and non-thermal plasma.}

\author[a]{Mohammad Shehzad Emritte,}
\author[a,1]{Sergio Colafrancesco,\note{Corresponding author.}}
\author[a]{Paolo Marchegiani}

\affiliation[a]{School of Physics, University of the Witwatersrand, Private Bag 3, WITS-2050, Johannesburg, South Africa}

\emailAdd{Sergio.Colafrancesco@wits.ac.za}
\emailAdd{emrittes@yahoo.com}
\emailAdd{Paolo.Marchegiani@wits.ac.za}

\abstract{Inverse Compton scattering of the anisotropic CMB fluctuations off cosmic electron plasmas generates a polarization of the associated Sunyaev-Zel'dovich (SZ) effect. The polarized SZ effect has important applications in cosmology and in astrophysics of galaxy clusters. However, this signal has been studied so far mostly in the non-relativistic regime which is valid only in the very low electron temperature limit for a thermal electron population and, as such, has limited astrophysical applications. Partial attempts to extend this calculation to the IC scattering of a thermal electron plasma in the relativistic regime have been done but these cannot be applied to a general relativistic or mildly relativistic electron distribution.
In this paper we derive a general form of the SZ effect polarization that is valid in the full relativistic approach for both thermal and non-thermal electron plasmas, as well as for a generic combination of various electron population which can be co-spatially distributed in the environments of galaxy clusters or radiogalaxy lobes. We derive the spectral shape of the Stokes parameters induced by the IC scattering of every CMB multipole for both thermal and non-thermal electron populations, focussing in particular on the CMB quadrupole and octupole that provide the largest detectable signals in cosmic structures (like galaxy clusters). We found that the CMB quadrupole induced Stoke parameter $Q$ is always positive with a maximum amplitude at a frequency of $\approx 216$ GHz which increases slightly with increasing cluster temperature. On the contrary, the CMB octupole induced $Q$ spectrum shows a cross-over frequency which depends on the cluster electron temperature in a linear way, while it shows a non-linear dependence on the minimum momentum $p_1$ of a non-thermal power-law spectrum as well as a linear dependence on the power-law spectral index of the non-thermal electron population. We discuss some of the possibilities to disentangle the quadrupole-induced $Q$ spectrum from the octupole-induced one which will allow to measure these important cosmological quantities through the SZ effect polarization at different cluster locations in the universe. We finally apply our model to the realistic case of the Bullet cluster and derive the visibility windows of the total, quandrupole-induced and octupole-induced Stoke parameter $Q$ in the frequency ranges accessible to SKA, ALMA, MILLIMETRON and CORE++ experiments.}

\begin{document}
 \maketitle
\section{Introduction}

When photons of the Cosmic Microwave Background (CMB) radiation pass through the atmosphere of a cosmic structure such as a galaxy cluster or a radio-galaxy, they are Comptonized by the electrons present. This causes a unique imprint on the intensity spectrum of the CMB which is usually referred to as the Sunyaev Zel'dovich  (SZ) effect \cite{Zeldovich1969, sunyaev1972, W1979, Birk1999}. The main component of this effect arises because of the scattering of the photons off a distribution of electrons which can be of thermal or non-thermal origin \cite{EK2000, CM2003}. Another component that gives rise to an SZ effect is the peculiar motion of the cosmic structure with respect to the CMB frame and this causes a kinematic spectral distortion (kSZ). The SZ/kSZ effect has been realized be a powerful probe in astrophysics since it can constrain the pressure, energetics  and spatial distributions of relativistic electrons in galaxy clusters and radio-galaxies \cite{CM2011radiogalaxies,CM2011, CM2015, EK2000}, test the acceleration history of cosmic rays during mergers \cite{Dogiel2007} and determine the nature of electron distributions \cite{Prokhorov2010, Prokhorov2011, Prokhorov2012}. In addition to that, the SZ/kSZ effect also has cosmological relevances for, e.g., independent determination of the Hubble constant \cite{Cavaliere1977, Silk1978, Birk1979}, revealing the nature of dark matter \cite{C2004, C2006}, constraining the equation of dark energy \cite{Weller2002}, probing the epoch of reionization (EOR) and the dark ages (DA) \cite{SKA2015, Cooray2006}. 
New physics can also be explored with the SZ effect such as the existence of massive photons \cite{CM2014} and non-Planckian effects in cosmological radio backgrounds \cite{CEM2015} due to the existence of a plasma frequency around the recombination epoch.

In addition to an intensity spectral distortion, the SZ effect also induces linear polarization in the CMB \cite{sazonov1999, lavaux2004}. The main source of polarization originates from the intrinsic multipoles of the CMB \cite{Cooray2003}, which exist due to the temperature variation of the CMB at the surface of last scattering created by spatial fluctuations in energy, bulk velocity and gravitational potential. Because of this variation in the CMB temperature, an anisotropic radiation field is seen by the electrons \cite{kolb1994} residing in cosmic structures and upon inverse Compton (IC) scattering a SZ effect polarization is produced. Using a non-relativistic approach, it has been shown that the SZ effect polarization depends only on the quadrupole of the CMB at the cluster's redshift \cite{sazonov1999, lavaux2004} and is of the order of $a_{2,2} \tau$, where $a_{2,2}$ is the temperature quadrupole of the CMB and $\tau$ is the optical depth of the electrons residing in the cosmic structure.

The quadrupole of the CMB is subjected to cosmic variance when measured from only our location since we observe only one sphere of the last scattering surface. The SZ effect polarization gives us, on the contrary, access to other spheres of the last scattering since it allows one to measure the quadrupole as seen from other locations in the Universe where cosmic structures are located. It has been discussed in \cite{kamion1997, P2004} that this measurement has the potential of reducing the cosmic variance on the CMB quadrupole by measuring the latter at several galaxy cluster locations. Furthermore, it has been also pointed out that the SZ effect polarization has  the power to test the homogeneity of the universe \cite{Maartens2011}. By determining the quadrupole of the CMB at other places in the Universe, it also tells us about the isotropy of the CMB at these places \cite{Maartens2011}. Since homogeneity cannot be measured directly, one can link it to isotropy via the Copernican principle (CP), i.e. that there is no special position in the Universe, and hence test for its validity. Also using polarization data from a sample of clusters over a wide range in redshift, the presence of the Integrated Sachs-Wolfe (ISW) \cite{Sachs1967} effect can be statistically established and its redshift dependence contribution to the rms quadrupole can be determined. Given the strong dependence of the ISW effect on the background cosmology, the cluster polarization can eventually be used as a probe of the dark energy \cite{Cooray2003}.

The non-relativistic SZ effect polarization signal for a cosmic structure with optical depth $\tau \approx 0.02$ is expected to be $\sim 0.1 \mu$K in the Rayleigh-Jeans frequencies \cite{sazonov1999, lavaux2004}, which is still below the detection limit of current instruments. One way to overcome this limitation has been commented in \cite{sazonov1999, lavaux2004}, that an r.m.s value of the quadrupole-induced SZ polarization can be established if the signal is measured for a large number of clusters.

Furthermore better experimental opportunities are foreseen with upcoming instruments such as the SKA \cite{Carilli2008, Dewney2013}, which will operate in the frequency range 0.03 GHz up to 40 GHz and whose sensitivities are around $\mu$Jy level, the ground-based ALMA experiment \cite{Carilli2008}, the space-borne MILLIMETRON experiment \cite{millimetron}, operating in the millimeter frequency range of 84-720 GHz and 100-1900 GHz respectively and the CORE++ space-borne survey experiment \cite{CORE++}; the combination of these experiments will provide a multifrequency spectral approach for detecting the SZ polarization signal. Given the coming experimental opportunities, it has become relevant to study the SZ effect polarization in depth, analyzing the possible astrophysical and cosmological aspects and their exploitation in the light of the achievable experimental sensitivities,

From a theoretical perspective, most of the previous works on SZ effect polarization have been performed in the non-relativistic regime which is valid only in the very low electron temperature limit for a thermal electron distribution \cite{sazonov1999, lavaux2004}. However, recent observations of galaxy clusters have revealed that the temperature of the intracluster medium (ICM) hosted by some of these cosmic structures can reach up to 14 keV on average \cite{Wik2014} and even up to 20 keV in some cluster regions \cite{Gomez2004}. At these temperatures, relativistic effects become important and if one wants to use the cluster SZ effect polarization for cosmological purposes, these relativistic effects have to be necessarily considered appropriately.

A previous work on SZ effect polarization \cite{CL1998} in the relativistic regime suggested that higher multipoles of the CMB can contribute to the polarization caused by the IC scattering process. This would allow to probe not only the CMB quadrupole, but also the CMB octupole at remote locations in the universe, allowing hence to reduce the cosmic variance on this higher multipole as well. The CMB octupole is important in probing homogeneity. Actually, it has been shown in \cite{Ellis1983, Maartens2011} that the vanishing of the CMB dipole, quadrupole and the octupole is a sufficient condition for a region to follow a Friedmann-Lemaitre-Robertson-Walker (FLRW) geometry. 

Furthermore, the detection of non-thermal emission (like, e.g., radio-halos) from galaxy clusters \cite{Feretti2012, Ferrari2008} and lobes of radio-galaxies also motivates for a full relativistic study of polarization in IC scattering processes. It is widely accepted that this non-thermal emission originates from a relativistic population of electrons spiraling around magnetic field lines. These non-thermal electrons \cite{Jaffe1977, Dennison1980} also contribute in the Comptonization of the CMB \cite{EK2000, CM2003} and hence an SZ effect polarization is also expected from them. It is therefore of interest to know whether the SZ effect polarization is coming from the thermal or the non-thermal electrons and evaluate the non-thermal effect in comparison with the thermal one. The SZ effect in intensity coming from the non-thermal electrons has been shown to extend from low to high frequencies, $\approx$ 1000 GHz. Based on this evidence, it can be anticipated that the SZ effect polarization spectrum will also span over a wide range of frequencies. This gives rise to an opportunity for searching the SZ effect polarization at frequencies around 100 GHz to 1000 GHz.

Matters are more complicated if the non-thermal emission regions co-spatially exist with the thermal X-ray emitting regions in galaxy clusters \cite{CM2003, CM2011}. In addition, it is also possible to have two or more thermal electron distributions, with different optical depth and temperature, co-existing together. It has been shown in \cite{CM2003} how to compute the SZ effect intensity spectrum for a general combination of various electron populations. By applying this technique to the SZ data observed for the Bullet cluster \cite{CM2011, CM2015}, it has been found that  the fit to the data is improved by using a combination of electron populations. Therefore it is important to extend this technique and compute the SZ effect polarization for a combination of electron populations.

We note that the relativistic derivation of the SZ effect polarization by \cite{CL1998, Itoh1998} has been done only in the case of a thermal electron distribution by expanding the relativistic Boltzmann equation in terms of the electron temperature parameter $\theta_e ={k T_e}/{m_e c^2}$. The approach used by these authors does not apply to a general electron distribution and the restriction to thermal electrons is somehow incomplete; therefore a more complete solution is needed for general electron distributions found in various astrophysical plasma, and in particular for power-law electron distributions which are present in cluster containing radio-halos and in the lobes/jets of radio galaxies.

We also note that the spectral features of the CMB octupole-induced SZ effect polarization have not been calculated so far, while this is an important task if one wants to disentangle the CMB octupole from the quadrupole term. Also the formalism that has been previously presented is somehow cumbersome, and this comparison is discussed in \cite{PB1}. Finally, no extended comparison with sensitivities of current or future instruments has been discussed so far, while this is a relevant issue for the observability of this effect.

In this work we compute the SZ effect polarization by solving the polarized relativistic Boltzmann equation given in \cite{PB1}. We extract the Stokes parameters and compute their spectrum for the quadrupole and the octupole of the CMB in the case of both thermal and non-thermal population of electrons. This approach also allows us to compute the polarization signal arising from a general combination of various electron populations. In order to assess the detectability of the signal, we also compute the expected signals for a real cluster like the Bullet cluster and we compare it with the sensitivity of various instruments operating in different frequency bands.

We mention here that there are secondary sources of SZ effect polarization \cite{sazonov1999, lavaux2004} other than those generated by the intrinsic multipoles of the CMB. The first one to mention is the polarization induced by the transverse motion of the cosmic structure with respect to the CMB frame. The amplitude of this effect is of the order of $(V_T/c)^2 \tau$, where $c$ is the speed of light and $V_T$ is the component of the peculiar velocity perpendicular to the line of sight.

Another source of polarization is that induced by multiple scattering, which is therefore of the order of $\tau^2$. 
The relativistic effects to the SZ effect polarizations induced via kinematic and multiple scattering can have important astrophysical relevance and will be considered in details in a forthcoming paper.

The structure of the paper is the following: in Sect. 2 we start our analysis from the known derivation of the SZ effect polarization in the non-relativistic case in order to show the links with the full relativistic derivation that will be discussed in Sect.3. The redistribution functions of polarized photons will be computed for thermal and non-thermal electron distributions, and the full relativistic Stokes parameters will be derived. We will then apply in Sect.4 our predictions to the realistic case of the Bullet cluster which hosts a superposition of thermal and non-thermal plasma.
We will finally discuss our results and present our conclusions in the final Sect.5.

Throughout the paper, we use a flat, vacuum--dominated cosmological model with $\Omega_m = 0.308$, $\Omega_{\Lambda} = 0.692$ and $H_0 =67.8$ km s$^{-1}$ Mpc$^{-1}$ \cite{Ade2015}.

\section{SZ effect polarization: non-relativistic regime}

For the sake of clarity, before computing the SZ effect polarization in a full relativistic approach, we first derive the SZ effect polarization in the non-relativistic case and then, in the next section, we describe the extension of this derivation to the relativistic domain.
 
\begin{figure}
\centering
\includegraphics[scale= 0.5,width=100mm,height=80mm]{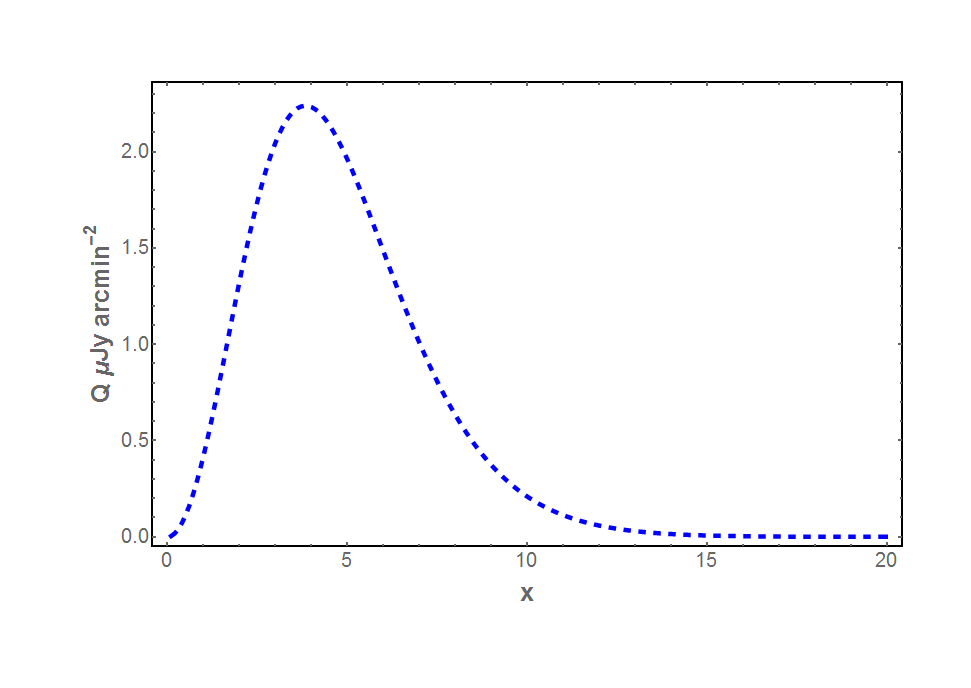}
\caption{The Stokes parameter $Q$ computed in the non-relativistic approach for a value $| a_{2,2}| =1.3 \times 10^{-5}$ and $\tau = 0.01$.}
\label{Qnr}
\end{figure}

A first approach in the study of the SZ effect polarization is based on the assumptions that the speed, $v_e$, of the electrons which scatter the CMB photons is small, i.e. $ \beta_{e}=v_e/c << 1 $, and that the Thomson limit is valid, that is $ h\nu << m_{e} c^{2}$. Under the second assumption, the process can be described using Thomson scattering. Assuming that the incident radiation is not polarized but anisotropic, the outgoing radiation will have a degree of linear polarization proportional to the CMB quadrupole moment in the angular distribution of the incident radiation. Choosing a frame of reference in such a way that the Z-axis coincides with the line of sight of the scattered radiation at first scattering, the Stokes parameters $Q$ and $ U $ are given by the following integral \cite{Chandra1960}:
\begin{equation}
\frac{\partial Q}{\partial \tau} (x)=\frac{3}{16 \pi} \int  \sin^2 (\theta) \cos (2 \phi) I (x,\theta,\phi) d \Omega 
\label{Qnr1}
\end{equation}
\begin{equation}
\frac{\partial U}{\partial \tau} (x) = \frac{3}{16 \pi} \int \sin^2 (\theta) \sin (2 \phi) I ( x, \theta, \phi) d \Omega \;,
\label{Unr1}
\end{equation}
where $x={h \nu}/{k T_0}$ and $T_0=2.725$ K is the average temperature of the CMB. The angle $\theta$ is the polar angle measured with respect to the Z-axis whereas $\phi$ is the azimuth angle. The intrinsic anisotropy of the incoming radiation in the case of the CMB is given by the primordial fluctuations of the temperature dependent unit vector $ \hat{n} ( \theta,\phi )$. Thus $I (x, \theta, \phi )$  is written as:
\begin{equation}
I (x,\theta,\phi ) = 2 \frac{\big(k T_0 \big)^3}{\big(h c \big)^2} \frac{x^{3}}{\exp \left[\frac{h \nu }{k T ( \theta, \phi )}\right] -1}=\sum_{l,m}^\infty I_{l,m}(x) Y_{l,m}(\theta, \phi) \;,
\label{Iani}
\end{equation}
where $T(\hat {n})$ is given by
\begin{equation}
T(\hat{n})= T_{0} [ 1 + \delta ( \theta, \phi ) ]
\label{Tvar}
\end{equation}
and
\begin{equation}
\delta (\theta, \phi ) = \sum_{l,m}^{\infty} a_{l,m} Y_{l,m}(\theta, \phi ) \;.
\label{delta}
\end{equation}
By inserting eq. \ref{delta} into eq. \ref{Tvar} and then substituting into eq. \ref{Iani}, we can write the intensity of the incident radiation as an expansion in terms of the spherical harmonics, given that the variations in the temperature of the CMB are generally very small:
\begin{equation}
 I ( x, \theta, \phi ) =\frac{ 2 (k T_{0})^3 }{(h c)^2}\bigg[ \frac{x^3}{e^{x}-1} + \frac{e^x x^{4}}{(e^x-1)^2} \sum_{l,m}^{\infty} a_{l,m} Y_{l,m} ( \theta, \phi ) \bigg] + O (\delta^2)=\sum_{l,m}^\infty I_{l,m}(x) Y_{l,m}(\theta, \phi) \;.
\label{Intensityexpansion}
\end{equation}
After inserting this into eq. \ref{Qnr1} and eq. \ref{Unr1} and integrating over the solid angle, we are left with only two terms, namely $ l=2, m=\pm 2$. The solution can be written as
\begin{equation}
\frac{\partial Q}{\partial \tau} (x)=\sqrt{\frac{3}{10 \pi}} \frac{I_{2,2} + I_{2,-2}}{4}={1 \over 2} \sqrt{3 \over{10 \pi}} Re[ I_{2,2} (x)] \;,
\label{Qint1}
\end{equation}
and
\begin{equation}
\frac{\partial U}{\partial \tau} (x)= \sqrt{3 \over {10 \pi}} \frac{I_{2,-2} + I_{2,-2}}{4 i}=-{1 \over 2} \sqrt{3 \over{10 \pi}} Im[ I_{2,2} (x)]  \;.
\label{Uint1}
\end{equation}
\newline
The multipoles of the intensity can be obtained directly from eq. \ref{Intensityexpansion} and the relevant ones up to the octupole are 
\begin{eqnarray}
&& I_{0,0}(x) = \sqrt{4\pi}\ 2\frac{(k T_0)^3}{(h c)^2} \frac{x^3}{e^x-1}=\ \sqrt{4\pi}\ 2\frac{(k T_0)^3}{(h c)^2} F_0 (x) \nonumber \\
&& I_{2,2}(x) = a_{2,2}\ 2\frac{(k T_0)^3}{(h c)^2}\frac{e^x x^4}{(e^x - 1)^2}=  \ a_{2,2}\ 2\frac{(k T_0)^3}{(h c)^2}F_1 (x)\nonumber \\
&& I_{3,2}(x) = a_{3,2}\ 2\frac{(k T_0)^3}{(h c)^2}\frac{e^x x^4}{(e^x - 1)^2}= \ a_{3,2}\ 2\frac{(k T_0)^3}{(h c)^2} F_1 (x) \;,
\label{intensitymultipoles}
\end{eqnarray}
where we have defined the functions $F_0(x)=x^3/(e^x-1)$ and $F_1(x)=(e^x x^4)/(e^x-1)^2$.
We have also used here the fact that $I^*_{l,m}={(-1)^m} I_{l,-m}$. Then we obtain the Stokes parameter $Q$ and $U$ as follows:
\begin{equation}
\frac{\partial Q}{\partial \tau} (x)= {1 \over 2} \sqrt \frac{3}{10 \pi}\ 2\frac{(k T_{0})^3}{(h c)^2} Re [ a_{2,2} ] F_{1}(x) \;,
\label{dQtaunr}
\end{equation}
and 
\begin{equation}
\frac{\partial U}{\partial \tau} (x)=-{ 1 \over 2 }\sqrt \frac{3}{10 \pi}\ 2\frac{(k T_{0})^3}{(h c)^2} Im [a_{2,2}] F_{1} (x) \;.
\label{dUtaunr}
\end{equation}
The Stokes parameters can be obtained in terms of the optical depth of the electron distribution in the single scattering approximation by just multiplying by $\tau$ as follows
\begin{equation}
Q (x)= {\tau \over 2} \sqrt \frac{3}{10 \pi}\ 2\frac{(k T_{0})^3}{(h c)^2} Re [ a_{2,2} ] F_{1}(x) \;,
\label{Qtaunr}
\end{equation}
and 
\begin{equation}
U (x)=-{ \tau \over 2 }\sqrt \frac{3}{10 \pi}\ 2\frac{(k T_{0})^3}{(h c)^2} Im [a_{2,2}] F_{1} (x) \;.
\label{Utaunr}
\end{equation}
The basis used to describe the Stokes parameters can always be rotated in such a way that $Re[a_{2,2}]=|a_{2,2}|$ and $Re[a_{3,2}]=|a_{3,2}|$. Hence we speak only of $Q$ as $U$ will be zero using such a basis. 
The values of $a_{l,m}$ are related to the coefficients $C_{l}$ of the angular power spectrum of the CMB temperature anisotropy \cite{Planck2013cmbspectra}. One can write 
\begin{eqnarray}
|a_{2,2}| \approx \sqrt{C_2} \\
|a_{3,2}| \approx \sqrt{C_3} \;.
\label{Tempcoeff}
\end{eqnarray}
We obtained values of $|a_{2,2}|=1.3 \times 10^{-5}$ and $|a_{3,2}|=8.7 \times 10^{-6}$. We show in Fig. \ref{Qnr} the spectrum of the Stokes parameter $Q$ for the CMB quadrupole computed using eq. \ref{Qtaunr}. We finally define the degree of polarization as 
\begin{equation}
\Pi=\sqrt{Q^2+U^2}/I \;.
\label{degPol}
\end{equation}

\section{The polarized Boltzmann equation}

From now onwards we use the unit convention $c=1$ and $h=1$ except where otherwise specified. 
The covariant Boltzmann equation describes the Compton scattering of photons and electrons $ \gamma (\vec{p_1}) + e^{-} (\vec{q_1}) \longrightarrow \gamma (\vec{p_2}) + e^{-} (\vec{q_2}) $. In a lab-frame $ V_L^{\mu}=[1,0,0,0]$, the non-polarized equation is written as \cite{Itoh1998, NK2009}:
\begin{equation}
\frac{d f (\vec{p_1})}{d t}=2 \int d^{3} q_1  d^3 q_2  d^3  p_2  W \bigg[f (\vec{p_2}) g_e (\vec{q_2})-f (\vec{p_1}) g_e (\vec{q_1})\bigg] \;,
\label{nonpoleq}
\end{equation}
where the functions $f$ and $g_e$ are general functions describing the momentum distribution of the photons and electrons respectively, and $W$ is written as
\begin{eqnarray}
W&=&\frac{3 \sigma_T}{32 \pi} m_e^{2} \frac{X}{E_1  E_2  p_1  p_2} \delta^4 \big( p_1^{\mu} + q_1^{\mu} - p_2^{\mu}- q_2^{\mu} \big) \\
X&=&m_e^2 \bigg( \frac{1}{k_2} -\frac{1}{k_1} \bigg)^2 + 2 m_e \bigg( \frac{1}{k_1} - \frac{1}{k_2} \bigg) + \frac{1}{2} \bigg( \frac{k_1}{k_2} + \frac{k_2}{k_1} \bigg) \;,
\label{eq47}
\end{eqnarray}
and $k_1$ and $k_2$ are defined as follows:
\begin{eqnarray}
k_1=-p_1^{\mu} V_{2 \mu} \\
k_2=-p_2^{\mu} V_{2 \mu}  \;.
\label{eq48}
\end{eqnarray}
The quantity $V_{2\mu}$ is the four-velocity of the electron after collision. In the rest-frame $ V_L $, $\vec{p_1}$ and $\vec{p_2}$ represent the momentum of the photon before and after collision and  $\vec{q_1}$ and $\vec{q_2}$ represent the momentum of the electron before and after collision, respectively. The 4-vectors in the delta function are represented as $ p_1^{\mu}=\big( p_1,\vec{p_1} \big)$, $ q_1^{\mu}=\big( E_1, \vec{q_1} \big)$, $p_2^{\mu}=\big( p_2, \vec{p_2} \big) $ and $ q_2^{\mu} = \big( E_2, \vec{q_2} \big) $. The quantity $k_i$ represents the magnitude of the momentum of the photon with 4-momentum $p_i^{\mu}$ in the rest frame of $V_2$ where $i=1,2$. 
The time derivative ${d}/{d t}$ is 
\begin{equation}
\frac{d}{d t} = \frac{1}{p_1} p_1^{\alpha} \partial_{\alpha} \;.
\label{eq49}
\end{equation}
It is convenient to split the Boltzmann equation into two terms, i.e. "scattering in"  and "scattering out", of the momentum element $ d^{3} p_1$ which can be written as follows:
\newline
\begin{equation}
\frac{d f}{d t} = \frac{d f}{d t}_{in} - \frac{d f}{d t}_{out} \;.
\label{eq50}
\end{equation} 
The first term in this equation is the rate of scattering of photons with momentum $\vec{p_2}$ off electrons with momentum $\vec{q_2}$ into $d^3 p_1$ around $\vec{p_1}$, while the second term represents the rate of scattering of photons with momentum $\vec{p_1}$ off electrons $\vec{q_1}$ into $d^3 p_2$ around $\vec{p_2}$. We should also point out that this equation neglects stimulated emission as well as Pauli blocking but is still valid outside Thomson's regime where quantum effects are not negligible. 

We note that the Boltzmann equation for polarization exhibits the same features as the non-polarized equation except that the cross-section and the distribution functions become tensor quantities which in turn requires the use of projection tensors \cite{PB1}. The equation is written as follows:
\begin{eqnarray}
p_1 \frac{d f }{dt}^{\mu \nu} ( p_1^{m} , V_L^{m}) &  = & 
m_e^2 \sigma_T \int\frac{ d^{3} q_1}{E_1}\frac{ d^3 q_2}{E_2} \frac{d^3 p_2}{p_2} \delta^4 \big( p_1^{\mu} + q_1^{\mu} - p_2^{\mu} - q_2^{\mu} \big)  \nonumber \\
 & & \times P_{\alpha \beta}^{\mu \nu} ( p_1^{m} , V_L^{m} )   
\bigg[ \Phi_{\rho \sigma}^{\alpha \beta} (p_1^{m}, p_2^m, V_2^m) f^{\rho \sigma} ( p_2^m , V_L^m ) g_e (\vec{q_2})  \nonumber \\ 
& & - \phi^{\alpha \beta} ( p_1^m , V_L^m ) g_{\gamma \delta} \Phi_{\rho , \sigma}^{\gamma \delta} ( p_2^m , p_1^m , V_1^m ) f^{\rho \sigma} ( p_1^m , V_L^m ) g_e (\vec{q_1} ) \bigg], \nonumber \\
& &  
\label{delta1}
\end{eqnarray}
where $g_{\gamma \delta}=(-1,1,1,1)$ is the metric tensor.
Eq. \ref{delta1} is the relativistic polarized Boltzmann equation. We note that in the laboratory frame an observer is seeing the velocity of the electrons to be $V_1^m$ (before collision) and $V_2^m$ (after collision). The polarization tensor for photons with momentum $p_1^\mu$ for this observer is denoted by $f^{\mu \nu} ( p_1^m , V_L^m )$. The quantities $V_1^m$ and $V_2^m$ denote the 4-velocity of the electron before and after collision whose momentum is $q_1^m$ and $q_2^m$, respectively, whereas $ p_1^m$ and $p_2^m$ represent that of the photon before and after the interaction.

We clarify that writing the distribution function $f^{\mu \nu} ( p^m , V^m )$ does not mean that $f$ is a function of $ V^m$,  but is only a notation used to denote that $f$ is the distribution function of the observer traveling with velocity $V^m$; it also does not mean that we are evaluating $f^{\mu \nu}$ in his rest-frame. If one wants to obtain the distribution function in the rest-frame of the observer, one has to Lorentz-transform to the $V$ frame in order to do so. So the function 
$f^{\mu \nu} ( p^m , V^m )$ $\longrightarrow$ $f^{\mu \nu} ( p^0, \vec{p} , V^m )$ $\longrightarrow$ $f^{\mu \nu} ( \vec{p} , V^m )$   can be also written as $f^{\mu \nu} ( \vec{p} , V^m )$. 
Also for the scalar function the following relationship $f ( p^\mu )$ $\longrightarrow$ $f ( p^0 , \vec{p} )$ $\longrightarrow$ $f ( \vec{p} )$ holds.
The reason why we can write it in terms of only 3-vectors is because $ p^0 = | p |$ for the photon but it is also true for massive particles because $ p^0 = \sqrt{ p^2 + m^2} $. We also mention here that eq. \ref{delta1} can also be used to calculate the SZ effect polarization resulting from kinematic effects as well as from multiple scatterings effects. These two cases will be treated specifically in a forthcoming paper. 

The cross-section here becomes a tensor, as we mentioned previously, and is written as follows
\begin{eqnarray}
&&\Phi_{ m n }^{ \mu \nu } \big[ p_1^m , p_2^m , V_2^m \big] \longrightarrow \text{ is the scattering cross-section for $ ( p_2^m , V_2^m ) $ $\rightarrow$ $ p_1^m $ } \nonumber \\
&&\Phi_{ m n }^{ \mu \nu } \big[ p_2^m , p_1^m , V_1^m \big] \longrightarrow \text{ is the scattering cross-section for $ ( p_1^m , V_1^m ) $ $\rightarrow$ $ p_2^m $ }. \nonumber \\
\label{eq52}
\end{eqnarray}
The term $ \Phi^{\mu \nu}_{ m n }$  is an analogue of $X$ for the polarized case and is constructed out of projection tensors \cite{PB1} . The tensor $\phi^{\mu \nu }$ represents the normalized polarization tensor written as ${f^{\mu \nu }} / {f}$. 
Finally we have $ P_{\alpha \beta }^{ \mu \nu }$ which is constructed out of the projection tensors as follows:
\begin{equation}
P_{\alpha \beta }^{\mu \nu } ( p^m , V^m ) = P_{\alpha}^{\mu} ( p^m , V^m ) P_{\beta}^{\nu} ( p^m , V^m ) \;.
\label{eq53}
\end{equation}
This projection tensor $P_{\alpha \beta}^{\mu \nu} ( p_1^{m} , V_L^{m} )$   actually projects the terms in the right hand side of the polarized Boltzmann equation into the plane perpendicular to the photon with momentum $p_1^m$ and 4-velocity of the observer $ V_L^m$. In the rest frame of the observer $V_L^m$ the projection tensor has only spatial components \cite{PB1}. 
The cross-section term is written in terms of the projection tensors in the Thomson approximation as 
\begin{equation}
\Phi_{\gamma \delta}^{\mu \nu} ( p_2^m , p_1^m , V_1^m ) = \frac{3}{8 \pi} P_{\alpha \beta}^{\mu \nu} ( p_2^m , V_1^m ) P_{\gamma \delta}^{\alpha \beta} ( p_1^m , V_1^m ) \;.
\label{eq54}
\end{equation}
The $ \delta^4 ( p_1^\mu + q_1^\mu - p_2^\mu - q_2^\mu )$ can be integrated out by using the following relation:
\begin{equation}
\displaystyle{\frac{d^3 q_1}{ E_1 } = d^4 q_1^\mu \delta \big[ \frac{1}{2} ( q_1^\mu q_{1 \mu}  + m_e^2) \big]} \;.
\label{intdelta}
\end{equation} 
We also write the electron distribution function as $g_e (\vec{q}) = n_e f_e ( \vec{q} )$ where $ n_e$ is the electron number density.
We can also use the definition of optical depth, $ d \tau_e = n_e \sigma_T d t$, to get rid of the Thomson total cross-section.
 
The conservation of four-momentum equation is written as
\begin{equation}
\displaystyle q_1^m = q_2^m + p_2^m - p_1^m  \;.
\label{eq56}
\end{equation}
This acts as a constraint on $q_1^m$, and the delta function in eq. \ref{intdelta} can be simplified to 
\begin{equation}
\displaystyle{\delta \big[ \frac{1}{2} ( q_1^m q_{1 m} + m_e^2 ) \big] = \delta \big[ m_e \big( k_1 - ( k_2 + R_{12} ) \big) \big] \;.}
\label{eq57}
\end{equation}
We have also introduced a new variable, $ R_{12} = \displaystyle {{p_1^\mu p_{2 \mu}}/{ m_e }}$, which will be very useful for our following  calculations. 

Using all these simplifications we can now cast the Boltzmann polarized equation as follows
\begin{eqnarray}
p_1  \frac{\partial}{\partial \tau} f^{\mu \nu } ( p_1^m , V_L^m ) & = & m_e^2 \int \frac{ d^3 q_2}{ E_2 } \frac{ d^3 p_2 }{ p_2 } \delta \big[ m_e \big( k_1 - ( k_2 + R_{12} )\big) \big ]  \nonumber \\
& & \times P_{ \alpha \beta }^{ \mu \nu } ( p_1^m , V_L^m ) 
\bigg[ \Phi_{\rho \sigma}^{\alpha \beta } ( p_1^m , p_2^m , V_2^m ) f^{ \rho \sigma } ( p_2^m , V_L^m ) f_e ( \vec{q_2} )  \nonumber \\
& & - \phi^{\alpha \beta } ( p_1^m , V_L^m ) g_{\gamma \delta } \Phi_{\rho \sigma}^{ \gamma \delta } ( p_2^m , p_1^m ,V_1^m ) f^{\rho \sigma }( p_1^m , V_L^m ) f_e ( \vec{q_1} ) \bigg] \;. \nonumber \\
\label{deltaelimination}
\end{eqnarray}

\subsection{The distribution function in the Thomson approximation}

Now we use this formalism to derive the Stokes parameters of the scattered CMB radiation by an electron gas. To do this we will make three important assumptions which actually not only simplify the calculation but also allow our results to be cross-checked with those obtained in previous works \cite{CM2003, EK2000}. The three assumptions are:

1) Single Scattering approximation: this means that we assume that each photon is scattered once by the electrons. This is valid in the mild optical depth regime for the study of galaxy clusters (e.g $\tau \approx$ 0.01) and radio-galaxies (e.g $\tau \approx 1 \times 10^{-4}$).

2) Thomson's cross-section: this means that we are neglecting quantum effects and in this way the scattering in the electron rest-frame can easily be described by Thomson's scattering, which in turn simplifies the cross-section term. This is valid because the CMB photons are mostly found in the low frequency range of the electromagnetic spectrum, meaning that they are less energetic than the electrons residing in the cosmic structures we are interested in. We note that this approximation is valid for electrons with Lorentz factor less than $\gamma\sim10^{7}$ (see, e.g.,\cite{Fargion1998}).

3) Unpolarized incident CMB: what we mean by this assumption is that before scattering the CMB is completely unpolarized; even though this is not completely true, for most of our calculations it is a quite reasonable simplification because the degree of polarization of the CMB before collision is very small \cite{Kovac2002, Hu2003, Page2007}. 

With these assumptions in hand the polarized Boltzman equation can be simplified extensively. In addition to these assumptions we also make a small change in our notation, mainly $ q_2 \longrightarrow q_e $ and also $ V_2 \longrightarrow V_e $. 
For single scattering between CMB photons and electrons, the equation writes as
\begin{eqnarray}
p_1   \frac{\partial}{\partial \tau} f^{\mu \nu } ( p_1^m , V_L^m ) & = &
 m_e \int \frac{ d^3 q_e}{ \gamma_e } \frac{ d^3 p_2 }{ p_2 } \delta \big[ m_e \big( k_1 - ( k_2 + R_{12} )\big) \big]  \nonumber \\
& & \times P_{ \alpha \beta }^{ \mu \nu } ( p_1^m , V_L^m )
\bigg[ \Phi_{\rho \sigma}^{\alpha \beta } ( p_1^m , p_2^m , V_e^m ) f^{ \rho \sigma } ( p_2^m , V_L^m ) f_e ( \vec{q_e} )  \nonumber \\
& & - \phi^{\alpha \beta } ( p_1^m , V_L^m ) g_{\gamma \delta } \Phi_{\rho \sigma}^{ \gamma \delta } ( p_2^m , p_1^m ,V_1^m ) f^{\rho \sigma }
 ( p_1^m , V_L^m ) f_e ( \vec{q_1} ) \bigg] \;. \nonumber \\
\label{eq59}
\end{eqnarray}
\newline
Now we make use of our second assumption, i.e the Thomson limit, which writes as 
\begin{eqnarray}
\gamma_e \alpha_2 & << & 1 \nonumber \\
\alpha_2 & = & \displaystyle \frac{p_2}{m_e} \;.
\label{eq60}
\end{eqnarray}
\newline
We also use the cross-section that we introduced in the previous section written as
\begin{equation}
\Phi_{\gamma \delta}^{\mu \nu } ( p_k^m , p_i^m , V_i^m ) =\displaystyle{ \frac{3}{8 \pi}} P_{\alpha \beta}^{\mu \nu} ( p_k^m , V_i^m ) P_{ \gamma \delta }^{ \alpha \beta } ( p_i^m , V_i^m ) \;.
\label{eq61}
\end{equation}
One point to be noted here is that the projection tensors which project the distribution function perpendicular to $V_i^m$ and $p_i^m$ (where $ k,i=1,2$) are included in this cross-section term. Then we define the following  useful variables 
\begin{eqnarray}
&n_{12} =\displaystyle{ \frac{k_1}{p_1} }= \gamma_e \big( 1 - \vec{\beta_e}\cdot\hat{n_1}\big) \nonumber \\
&n_{22} = \displaystyle{\frac{k_2}{p_2}} = \gamma_e \big( 1 - \vec{\beta_e}\cdot \hat{n_2}\big) \nonumber \\
&r_{12} = \displaystyle{\frac{p_1^\mu p_{1 \mu}}{ p_1 p_2 }} = m_e R_{12} = \hat{n_1}\cdot\hat{n_2} - 1 \nonumber \\
&\alpha_j =\displaystyle{ \frac{ p_j}{m_e}}
\label{n12n22r12}
\end{eqnarray}
\newline 
where $\hat{n_1}$ and $\hat{n_2}$ are unit vectors in the direction of $\vec{p_1}$ and $\vec{p_2}$ and $\vec{\beta_e}$ is the electron velocity. 
The delta function $\delta \big[ m_e \big( k_1 - ( k_2 + R_{12}) \big) \big]$ can be further simplified by using the Thomson limit as follows:
\begin{eqnarray}
\displaystyle m_e \big[ k_1 - ( k_2 + R_{12} ) \big] & = & -{m_e^2} n_{22} \big[ \alpha_2 - \alpha_1 \frac{n_{12}}{n_{22}} (1 - \alpha_2 \frac{r_{12}}{n_{12}})\big] \nonumber \\
\displaystyle & = & -{m_e^2} n_{22} \big[ \alpha_2 - \alpha_1 \frac{n_{12}}{n_{22}} ( 1 - O ( \alpha_2 \gamma_e) ] \nonumber \\
\displaystyle &= &- {m_e^2} n_{22} \big[ \alpha_2 - \alpha_1 \frac{n_{12}}{n_{22}} \big] \;.
\label{eq63}
\end{eqnarray}
In order to arrive at the previous approximation we made use of the following inequality:
\begin{equation}
\displaystyle \alpha_{2} | \frac{ r_{12}}{ n_{12}} | \leq \frac{ 2 \alpha_2 }{ \gamma_e ( 1 - \beta_e ) }  
= 2 \alpha_2 ( 1 + \beta_e ) \gamma_e  \leq 4 \gamma_e \alpha_2 = O ( \gamma_e \alpha_2 ) \;. 
\label{eq64}
\end{equation}
In the Thomson limit, and in the rest frame of the electrons the magnitude of the momentum of the photon before and after collision is the same, hence  $k_1 \approx k_2$, and therefore the variable $p_2$ is constrained by the following condition:
\begin{equation}
\displaystyle p_2 = \frac{n_{12}}{n_{22}} p_1 \;.
\label{eq65}
\end{equation}
Another simplification can be made by noticing that
\begin{equation}
\gamma_1 = \gamma_e \big[ 1 + O ( \alpha_2 \gamma_e ) \big] \;.
\label{eq66}
\end{equation}
This can be achieved by putting $ m = 0 $ in the equation $q_1^m = q_2^m + p_2^m - p_1^m$. Using $\alpha_1 =\big( {n_{22}}/{n_{12}}\big)\ \alpha_2$ we obtain
\begin{equation}
\gamma_1 = \gamma_e + \alpha_2 \left[ 1 - \frac{n_{22}}{n_{12}}\right] \;.
\label{eq67}
\end{equation}
Then one can show that:
\begin{eqnarray}
\gamma_1 & = & \gamma_e \bigg( 1 + \frac{\alpha_2}{\gamma_e}\ \big( 1 - \frac{n_{22}}{n_{12}}\big)\bigg) \leq \nonumber \\
& & \leq  \gamma_e \bigg( 1 + \frac{\alpha_2}{ \gamma_e }\ \bigg| 1 - \frac{n_{22}}{n_{12}} \bigg| \bigg) = \nonumber \\
& & =  \gamma_e \bigg[ 1 + 2 \beta_e \alpha_2 \big( 1 + \beta_e \big) \gamma_e \bigg] \leq \nonumber \\
& & \leq \gamma_e \bigg[ 1 + 4 \alpha_2 \gamma_e \bigg] = \nonumber \\
& & = \gamma_e \bigg[ 1 + O ( \alpha_2 \gamma_e ) \bigg] \;.
\label{gamma1=gamma2}
\end{eqnarray}
To arrive at the result we use the following inequalities:
\newline
\begin{eqnarray}
\bigg| \alpha_2 \big( 1 - \frac{n_{22}}{n_{12}} \big) \bigg| & \leq & \bigg| \alpha_2 \bigg[ 1 - \frac{1+ \beta_e }{1-\beta_e} \bigg] \bigg| = \nonumber \\
& & =  \alpha_2 \bigg| \frac{ -2 \beta_e}{ 1 - \beta_e } \bigg| = \nonumber \\
& & =  2 \beta_e \alpha_2 \big( 1 + \beta_e \big) \gamma_e^2 \leq \nonumber \\
& & \leq 4 \alpha_2 \gamma_e^2 \;.
\label{eq69}
\end{eqnarray} 
\newline
From eq. \ref{gamma1=gamma2}, it can be noticed that the energy of the electrons is unaltered during the Thomson scattering, hence also the distribution function
\newline
\begin{equation}
f_e ( \vec{q_1} ) \approx f_e ( \vec{ q_e} ) \;.
\label{eq70}
\end{equation}
\newline 
Using these simplifications we arrive at the following expression:
\begin{eqnarray}
\displaystyle \frac{\partial}{\partial \tau} f^{\mu \nu } ( p_1^m , V_L^m ) & = & \frac{3}{8 \pi} \int \frac{d^3 q_e}{ \gamma_e} \int d \Omega_2 \displaystyle \frac{ n_{12}}{ n_{22}^2} f_e ( \vec{q_e} )  \nonumber \\
& & \times \bigg[ J_{\alpha}^{\mu} ( p_1^m , V_e^m , V_L^m ) J_{\beta}^{\nu} ( p_1^m , V_e^m , V_L^m ) f^{ \alpha \beta } ( p_2^m , V_e^m )  \nonumber \\
& & - \phi^{ \mu \nu } ( p_1^m , V_L^m ) P_{ \alpha \beta } ( p_2^m , V_1^m ) f^{ \alpha \beta } ( p_1^m , V_1^m ) \bigg] \;,
\label{2ndassump}
\end{eqnarray}
with
\begin{equation}
\displaystyle J_{\alpha}^{\mu} ( p_1^m , V_e^m , V_L^m ) = P_{\beta}^{\mu} ( p_1^m , V_L^m ) P_{\alpha}^{\beta} ( p_1^m , V_e^m ) \;.
\label{Jp1}
\end{equation}
\newline
Now we make use of the third assumption which is that the CMB is unpolarized prior to the scattering by the electrons. With this assumption one can make the following replacements
\begin{eqnarray}
f^{\mu \nu} ( p^m , V^m ) & = & \frac{1}{2} f ({ p^m }) P^{\mu \nu } ( p^m , V^m )\\
\phi^{\mu \nu} ( p^m , V^m ) & = & \frac{1}{2} P^{\mu \nu} ( p^m , V^m ) \;.
\label{eq72}
\end{eqnarray}
Finally eq. \ref{2ndassump} can be written as follows:
\begin{eqnarray}
 \displaystyle \frac{\partial}{\partial \tau} f^{\mu \nu} ( p_1^m , V_L^m ) & = & \frac{3}{16 \pi} \int \frac{d^3 q_e}{ \gamma_e } \int d\Omega_2 \frac{ n_{12}}{ n_{22}^2} f_e ( \vec{q_e} )  \nonumber \\
& & \times \displaystyle \Bigg[ \bigg[ P^{\mu \nu} ( p_1^m , V_L^m ) - L^{\mu} L^{\nu} ( p_1^m , p_2^m , V_e^m ) \bigg] f ( p_2^m)  \nonumber \\
& & - P^{ \mu \nu } ( p_1^m , V_L^m ) \bigg[ 1 + \eta_{12} \big( 1 + \frac{1}{2} \eta_{12} \big) \bigg] f ( p_1^m ) \Bigg] \;,
\label{Last}
\end{eqnarray}
where we define
\begin{eqnarray}
\displaystyle L^{\mu} ( p_1^m , p_2^m , V_e^m ) & = & \frac{1}{n_{22}} \bigg[ \frac{ p_2^\mu}{ p_2} - \bigg( 1 + \gamma_e \frac{r_{12}}{n_{12}} \bigg) \frac{p_1^\mu}{ p_1 } + \frac{r_{12}}{n_{12}} V_e^\mu \bigg] \nonumber \\
\displaystyle \eta_{12} & = & \frac{r_{12}}{n_{12} n_{22}} \nonumber \\
\displaystyle L^{\mu} L_{\mu} & = & - 2 \eta_{12} \bigg( 1 + \frac{1}{2} \eta_{12} \bigg) \;.
\label{eq74}
\end{eqnarray}
In these last equations we recall that $p_1^\mu$ or $p_2^\mu$ is the momentum in the frame $V_L^\mu = \big[ 1, 0 ,0 ,0 \big]$ and from this position we can write that $ p_k = - p_k^\mu V_{L \mu}$. Since we are using $c=1$ and $h=1$, then $p_1$ and $\nu_1$ can be interchanged at will.

\subsection{Stokes parameters}

We show here how the Stokes parameters are derived from the tensor $f^{\mu \nu} ( p_1^m , V_L^m )$. 
We first derive the first Stokes parameter $ I $ which is given by:
\begin{eqnarray}
\displaystyle \frac{\partial}{\partial \tau} I ( \vec{p_1} ) = p_1^3 \frac{d}{d \tau} f_{\mu}^{\mu} ( \vec{p_1} ) & = & \frac{3 p_1^3}{8 \pi} \int \frac{d^3 q_e}{ \gamma_e} \int d \Omega_2 \frac{n_{12}}{n_{22}^2} f_e ( \vec{q_e} )  \nonumber \\
&& \times \bigg[ 1 + \eta_{12} \bigg( 1 + \frac{\eta_{12}}{2} \bigg)\bigg] \big[ f (\vec{p_2} ) - f ( \vec{p_1} ) \big] \;, \nonumber \\
\label{I11}
\end{eqnarray}
where we have used the notation $f ( p^\mu ) = f ( \vec{p} )$ that we already discussed before, and the fact that $ I = p^3 f$ which is the relation between the distribution function of photons to the intensity. In order to determine the other Stokes parameters, namely $Q$ and $U$, the choice of basis matters here and, depending on how the basis are chosen, will determine the simplicity of the calculation \cite{PB1}. 
In our case we choose a system of basis perpendicular to the observed radiation, that is in usual term we choose our Z-axis to be along the direction of the observed radiation. In this way the tensor $ f^{\mu \nu } ( \vec{p_1} )$ can be written as follows:
\begin{eqnarray}
\displaystyle f^{ \mu \nu } ( \vec{p_1} ) =\displaystyle \frac{1}{2 p_1^3} \left[ \begin{array}{ c c c c} 0 & 0 & 0 & 0 \\ 0 & I ( \vec{p_1} ) + Q ( \vec{p_1} ) & U ( \vec{p_1} ) + i V ( \vec{p_1} ) & 0 \\ 0 & U ( \vec{p_1} ) - i V ( \vec{p_1} ) & I ( \vec{p_1} ) - Q ( \vec{p_1} ) & 0 \\ 0 & 0 & 0 & 0 \end{array}\right] \;.
\label{Matrix1}
\end{eqnarray}
We then extract the Stokes parameters from this matrix as follows:
\begin{eqnarray}
&&\frac{d}{d \tau} Q ( \nu_1 ) = \nu_1^3 \frac{d}{d \tau} \bigg[ f^{11} ( \nu_1 ) - f^{22} ( \nu_1 ) \bigg] \\
&&\frac{d}{d \tau} U (\nu_1 ) = \nu_1^3 \frac{d}{d \tau} \bigg[ f^{12} ( \nu_1 ) + f^{21} ( \nu_1 ) \bigg] \;.
\label{extractedQU}
\end{eqnarray}
In this coordinate system the following variables take the form:
\begin{eqnarray}
&&p_1^{\mu} = p_1 \big( 1 , 0 , 0 ,1 \big)  \\
&&r_{12} = \cos \theta_2 -1 \\
&&n_{12} = \gamma_e \big[ 1 - \beta_e \cos \theta_e \big] \\
&&n_{22} = \gamma_e \bigg[ 1 - \beta_e \big[ \cos \theta_2 \cos \theta_e + \sin \theta_2 \sin \theta_e \cos ( \phi_2 - \phi_e ) \big] \bigg] \;,
\label{variableform}
\end{eqnarray}
and also
\begin{eqnarray}
&&V_e^{\mu} = \gamma_e \bigg[ 1 , \beta_e \cos \phi_e \sin \theta_e , \beta_e \sin \phi_e \sin \theta_e , \beta_e \cos \theta_e \bigg] \\
&&p_2^{\mu} = p_2 \bigg[ 1 , \cos \phi_2 \sin \theta_2 , \sin \phi_2 \sin \theta_2 , \cos \theta_2 \bigg] \;.
\label{variableform2}
\end{eqnarray}
The Stokes parameters $Q$ and $U$ are then written as follows:
\begin{eqnarray}
\displaystyle \frac{\partial Q}{\partial \tau}(\nu_1) & = & -\frac{3}{16 \pi} \int \frac{d^3 q_e}{\gamma_e} \int d \Omega_2 \frac{I ( \nu_2, \hat{n}_2)}{n_{12}^4 n_{22}} f_e (\vec{q_e})  \nonumber \\
&&\displaystyle \times \big[ \cos 2\phi_2 \sin^2 \theta_2  n_{12}^2 + 2 \cos (\phi_2 + \phi_e ) \sin \theta_2 \sin \theta_e n_{12} r_{12} \gamma_e \beta_e   \nonumber \\
&&+ \displaystyle \cos 2\phi_e \sin^2 \theta_e r_{12}^2 \beta_e^2 \gamma_e^2 \big]   \\
\displaystyle \frac{\partial U}{\partial \tau}(\nu_1) & = & -\frac{3}{16 \pi} \int \frac{d^3 q_e}{\gamma_e} \int d \Omega_2 \frac{I ( \nu_2, \hat{n}_2)}{n_{12}^4 n_{22}} f_e (\vec{q_e})  \nonumber \\
&& \times \displaystyle \big[ \sin 2\phi_2 \sin^2 \theta_2  n_{12}^2 + 2 \sin (\phi_2 + \phi_e ) \sin \theta_2 \sin \theta_e n_{12} r_{12} \gamma_e \beta_e   \nonumber \\
&& + \displaystyle \sin 2\phi_e \sin^2 \theta_e r_{12}^2 \beta_e^2 \gamma_e^2 \big] \;.
\label{Stokesfull}
\end{eqnarray}
These 5-dimensional integrals can be evaluated by breaking them into five 1-dimensional integrals. We will compute them first for the intensity, and this will allow us to show that our result are consistent with the usual approach of computing the unpolarized SZ effect spectrum.

\subsection{The Stokes parameter $I$}

We compute here the integrals for the Stokes parameter $I$ first for an isotropic incident radiation and later we will introduce an anisotropy via a spherical harmonic expansion. We start by defining $ \mu_e = \cos \theta_e $, $\mu_2 = \cos \theta_2 $ and $ \phi_0 = \phi_2 -\phi_e$. 
With these new variables eq. \ref{I11} takes the following form:
\begin{eqnarray}
\displaystyle \frac{\partial f}{\partial \tau} ( \nu_1 ) & = & \frac{3}{32 \pi^2} \int_{0}^{1} d \beta_e \int_{-1}^{1} d \mu_e \int_{0}^{2 \pi} d \phi_e \int_{-1}^{1} d \mu_2 \int_{0}^{2 \pi} d \phi_0 \frac{n_{12}\ f_{e} (\beta_e)}{ \gamma_e n_{22}^2}  \nonumber \\
&& \displaystyle \times \bigg[1 + \eta_{12} \bigg(1+ \frac{\eta_{12}}{2} \bigg)\bigg] \bigg[ f ( \nu_2) -f ( \nu_1) \bigg] \;.
\label{eq81}
\end{eqnarray}
\newline
In order to evaluate the integrals above, we assume that in the laboratory-frame the electron plasma appears isotropic and hence the distribution function of the electrons becomes independent of directions and can be written in terms of $\beta_e$ as follows
\begin{equation}
\frac{1}{4 \pi} f_e ( \beta_e ) d \beta_e = f_e ( \vec{q_e} ) q_e^2 d q_e \;,
\label{eisotropy}
\end{equation}
\newline
and use the fact that for a generic function $\Psi (\phi_e , \phi_2 - \phi_e)$,
\begin{equation}
\displaystyle \int_{0}^{2 \pi} \int_{0}^{2 \pi} \Psi (\phi_e , \phi_2 - \phi_e)\ d \phi_2 \ d \phi_e = \int_{0}^{2 \pi} \int_{0}^{2 \pi} \Psi ( \phi_e , \phi_0 )\ d \phi_0\ d \phi_e \;.
\label{generic1}
\end{equation}
\newline
In order to further simplify eq. \ref{eq81} we also introduce another variable:
\begin{equation}
\chi_0 = \cos \phi_0 \;,
\label{eq84}
\end{equation}
and if we consider a general function $F ( \cos \phi_0 , \sin \phi_0 )$ which has trigonometric functions as its argument, then we can write
\begin{eqnarray}
\int_{0}^{2 \pi} F ( \cos \phi_0 , \sin \phi_0 ) d \phi_0 & = & 
 \int_{-1}^{1} \bigg[ F ( \cos \phi_0 \rightarrow \chi_0 , \sin \phi_0 \rightarrow \sqrt{ 1-\chi_0^2 })  \nonumber \\
&& +  F ( \cos \phi_0 \rightarrow \chi_0 , \sin \phi_0 \rightarrow -\sqrt{ 1-\chi_0^2 })\bigg] \frac{d\chi_0 }{\sqrt{1-\chi_0^2}} \;. \nonumber \\
& & 
\label{generic2}
\end{eqnarray}
By using the simplifications of eq. \ref{generic1} and \ref{generic2} and integrating eq \ref{eq81} over $ \phi_e $ we obtain:
\begin{eqnarray}
\frac{\partial f}{\partial \tau} ( \nu_1) & = & \frac{3}{16 \pi} \int_{0}^ {1} d \beta_e \int_{-1}^{1} d \mu_e \int_{-1}^{1} d \mu_2 \int_{-1}^{1} d \chi_0\ f_e ( \beta_e )  \nonumber\\
&& \times \displaystyle \frac{ 2 n_{12}^2 n_{22}^2 + 2 n_{12} n_{22} ( \mu_2 -1 ) + {(\mu_2-1)}^2}{ n_{12} n_{22}^4 \gamma_e \sqrt{1-\chi_0^2}} \big[ f (\nu_2 ) - f (\nu_1 ) \big] \;. \nonumber \\
& &
\label{fI}
\end{eqnarray}
At this stage we can perform a check that the polarized covariant Boltzmannn equation gives the same result as the Wright's method for the Stokes parameter $I$ by making a transformation into the electron frame \cite{NK2009} using the following variables:
\begin{eqnarray}
\displaystyle \mu_0 & =& \frac{\mu_2 -1}{n_{12} n_{22}} + 1 \nonumber\\
\displaystyle \mu & = & \frac{\gamma_e n_{12} -1}{n_{12} \gamma_e \beta_e}\\
\displaystyle \mu' & = & \frac{\gamma_e n_{22}-1}{n_{22} \gamma_e \beta_e} \;. \nonumber 
\label{wrights}
\end{eqnarray}
By doing that, we obtain an equation in terms of the new variables of the following form:
\begin{eqnarray}
\displaystyle \frac{\partial f}{\partial \tau} (\nu_1) & = & \frac{3}{16 \pi} \int d \beta_e \int d \mu \int d \mu' \int d \mu_0\big[ f \big( \nu_2 \big)-f\big(\nu_1\big)\big]f_e (\beta_e)  \nonumber \\
&& \times \displaystyle \frac{1+\mu_0^2}{\gamma_e^4 (1-\beta_e \mu)^3 \sqrt{ 1-\mu_0^2 -\mu^2-\mu'^{\ 2} + 2 \mu_0 \mu \mu'}} \;,
\label{pp1}
\end{eqnarray}
with 
\begin{eqnarray}
&&\mu_{0,min} =\mu \mu' - \sqrt{ (1-\mu^2) (1-\mu'^{\ 2})} \\
&&\mu_{0,max}=\mu \mu' + \sqrt{ (1-\mu^2) (1-\mu'^{\ 2})} \nonumber.
\label{restframevariable}
\end{eqnarray}
The integration over $\mu_0$ can be done easily and then introducing a last variable, which is related to the frequency shift 
\begin{eqnarray}
&& e^s= \frac{1-\beta_e \mu'}{1-\beta_e \mu}  \\
\nonumber \\ 
&&\displaystyle{ s=\ln (\nu_2/\nu_1)} \nonumber.
\label{shift}
\end{eqnarray}
The eq. \ref{pp1} can be cast into the usual form \cite{EK2000, CM2003, Birk1999} as follows
\begin{equation}
\displaystyle \frac{\partial f}{\partial \tau}(\nu_1)  =  \int_{-\infty}^{\infty} P_1 (s)\ \big[f (e^s \nu_1)-f(\nu_1)\big] ds \;,
\label{COOL}
\end{equation}
where 
\begin{eqnarray}
\displaystyle P_1(s) & = &\int_{\sinh {|s| \over 2}}^{1} f_e (p_e)\  P (s,p_e)\ d p_e \\
\displaystyle P(s,\beta_e) & = & \frac{3\ e^s}{32}\int_{\mu_{min}}^{\mu_{max}} \frac{(3-\mu^2) \beta_e^2 - (1-3 \mu^2) \big[ 1- e^s ( 1-\mu \beta_e)\big]}{\beta_e^3 \gamma_e^4 (1-\beta_e \mu)^2} \ d \mu \;. \nonumber
\label{cool}
\end{eqnarray}
The electron distribution function has been written in terms of the electron momentum $p_e$ and the function $P (s ,p_e)$ is just the function $ P (s,\beta_e)$ with the $\beta_e$ and $\gamma_e$ substituted in terms of $p_e$. This is given by
\begin{eqnarray}
&& \gamma_e=\sqrt{1+p_e^2} \nonumber \\
&& \beta_e = \frac{p_e}{\sqrt{1+p_e^2}} \;. 
\label{gammap}
\end{eqnarray}
The function $P_1(s)$ is same function used in \cite{EK2000, CM2003, Birk1999} and is normalized to one. This shows, comfortably, that our formalism is consistent with previous works. We are now in a position to include the distribution of CMB anisotropy in our treatment.

\subsection{Anisotropic incident CMB radiation}

\begin{figure}
\begin{center}
\begin{tabular}{c c}
\includegraphics[width=80mm,height=75mm]{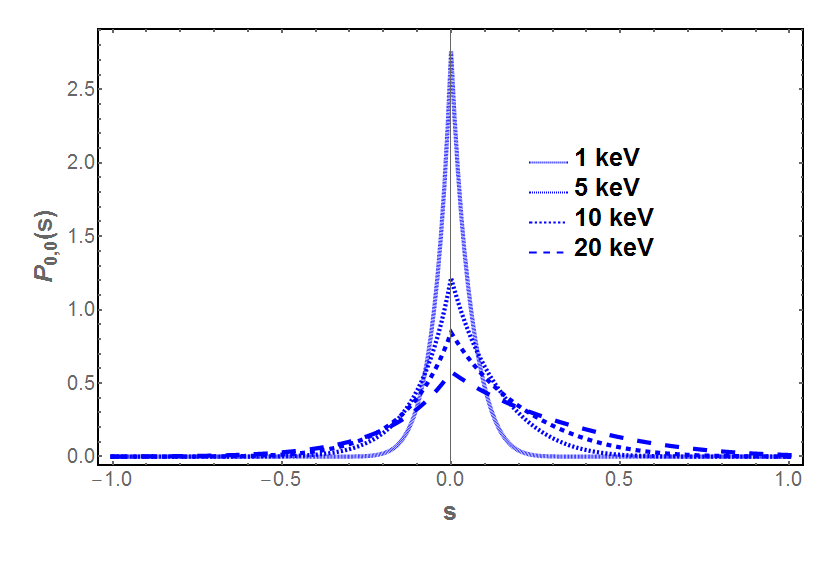} & \includegraphics[width=80mm,height=75mm]{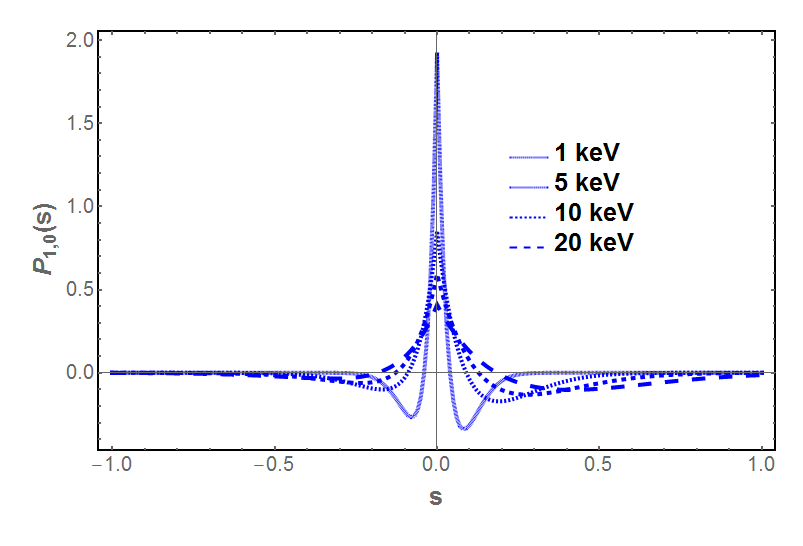}\\
\includegraphics[width=80mm,height=75mm]{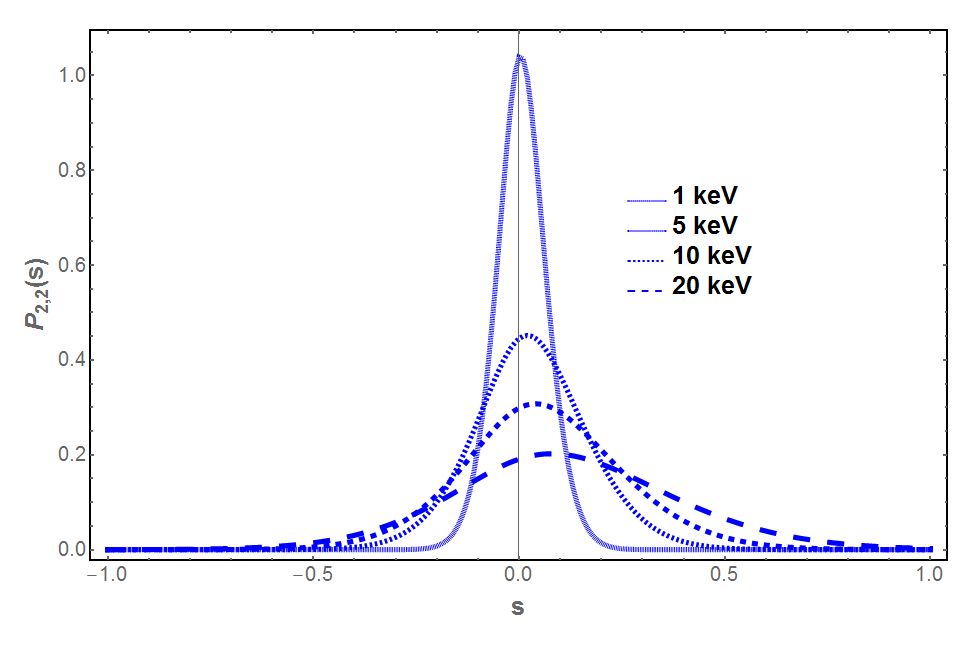} & \includegraphics[width=80mm,height=75mm]{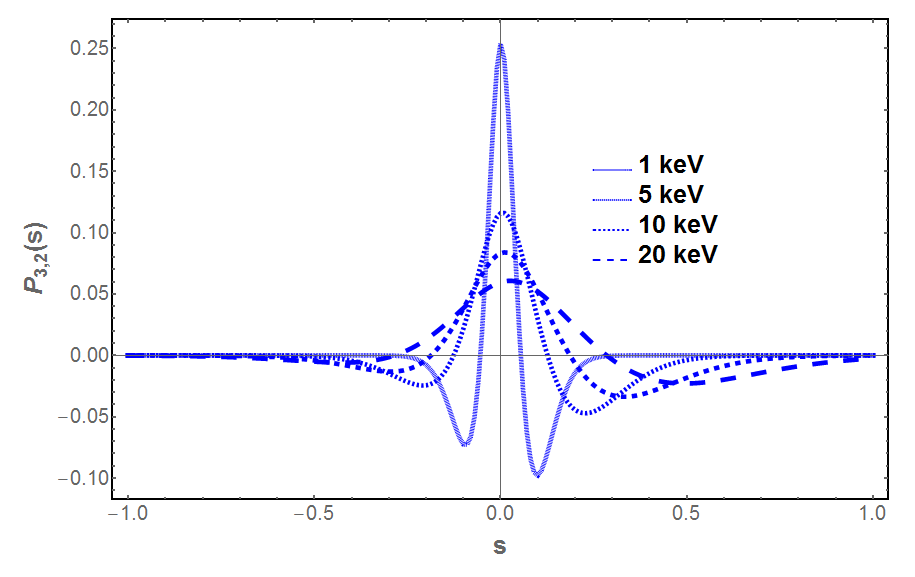}
\end{tabular}
\caption{The redistribution function $P_{l,m}(s)$, for $l=0,1,2,3$ and $m=0,2$, for thermal electrons at different temperatures as indicated.}
\label{redistributionfunctionth}
\end{center}
\end{figure}

\begin{figure}
\begin{center}
\begin{tabular}{c c}
\includegraphics[width=80mm,height=75mm]{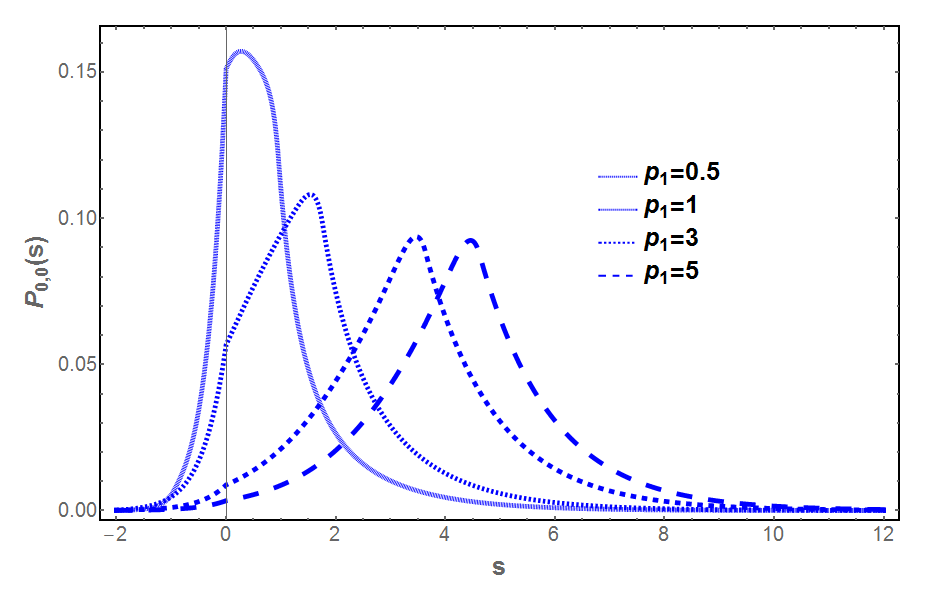} & \includegraphics[width=80mm,height=75mm]{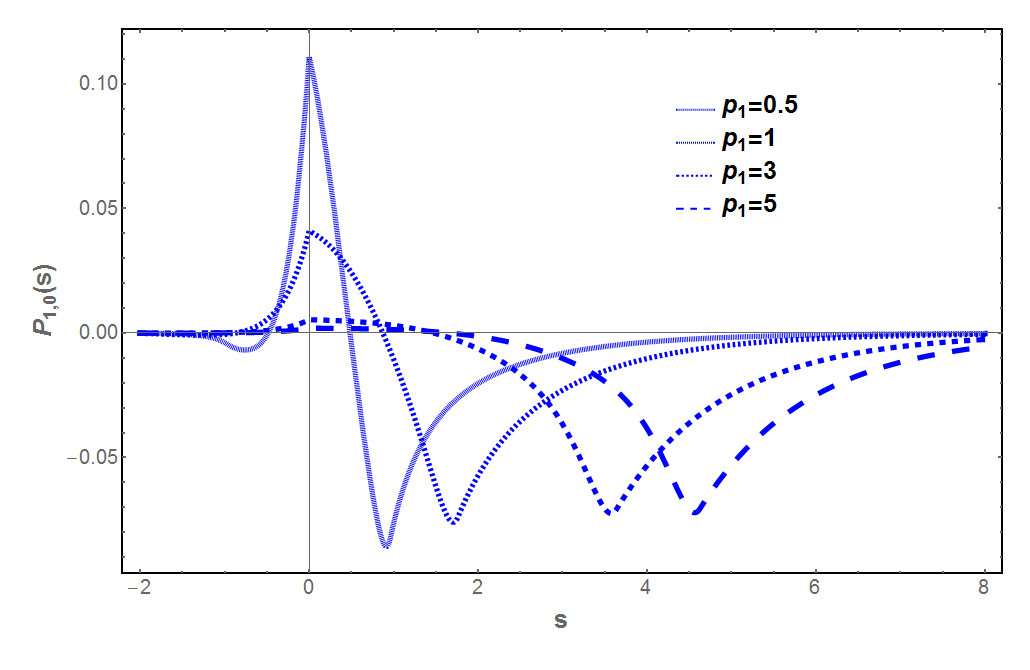}\\
\includegraphics[width=80mm,height=75mm]{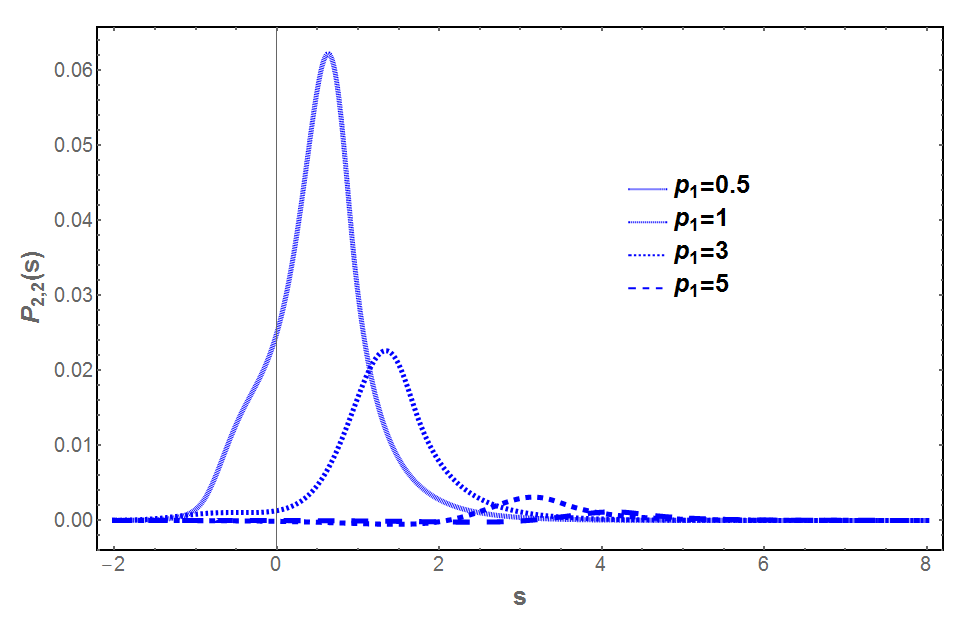} & \includegraphics[width=80mm,height=75mm]{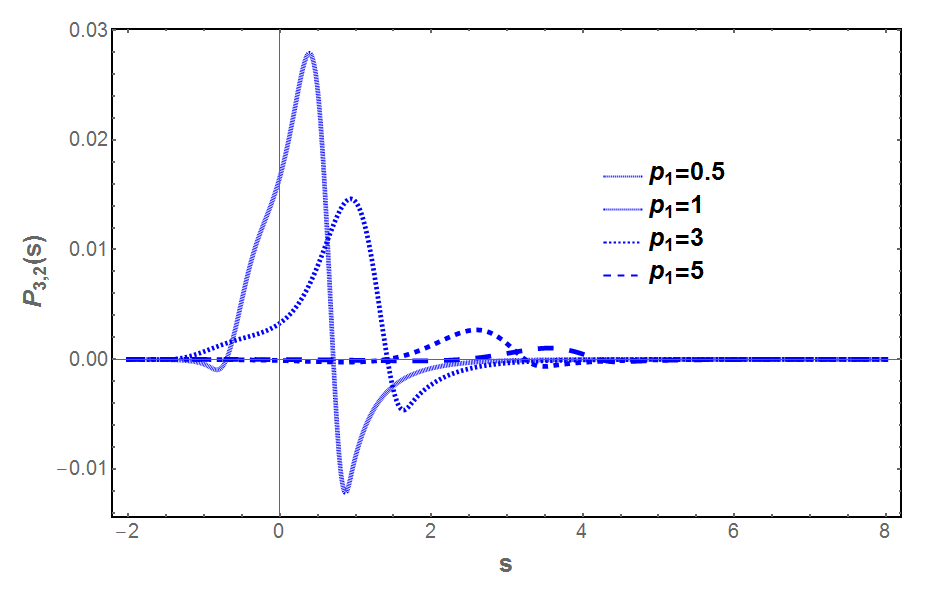}
\end{tabular}
\caption{The redistribution function $P_{l,m}(s)$, for $l=0,1,2,3$ and $m=0,2$, for a single power law distribution of electrons with spectral index $\alpha=2.5$ for different minimum momenta $p_1$.}
\label{redistributionfunctionspl}
\end{center}
\end{figure}

The rate of change of the distribution function can be broken down into two terms
\begin{equation}
\frac{\partial f}{\partial \tau} (\nu_1,\hat{z})= \frac{\partial f}{\partial \tau} \bigg|_{in} (\nu_1,\hat{z}) - \frac{\partial f}{\partial \tau} \bigg|_{out} (\nu_1,\hat{z}) \;.
\label{eq92}
\end{equation}
The rate of ''scattering out'' can easily be integrated (right-hand side of eq. \ref{fI} given the fact that $P_1(s)$ is normalized to one)  and the result is:
\begin{eqnarray}
\frac{\partial f}{\partial \tau} ( \nu_1,\hat{z}) & = & \frac{3}{16 \pi} \int_{0}^ {1} d \beta_e \int_{-1}^{1} d \mu_e \int_{-1}^{1} d \mu_2 \int_{-1}^{1} d \chi_0\ f_e ( \beta_e )  \nonumber\\
&& \times \displaystyle \frac{ 2 n_{12}^2 n_{22}^2 + 2 n_{12} n_{22} ( \mu_2 -1 ) + {(\mu_2-1)}^2}{ n_{12} n_{22}^4 \gamma_e \sqrt{1-\chi_0^2}} f (\nu_1,\hat{z} ) \nonumber \\
&&=\int_{-\infty}^{\infty} P_{1}(s) f(\nu_1, \hat{z}) \ ds = 1
\label{fout}
\end{eqnarray}
To determine the ''scattering in'' we expand the distribution function in a spherical harmonic series as follows:
\begin{eqnarray}
f (\nu_1, \hat{n}) & = & \sum_{l=0}^{\infty} \sum_{m=-l}^{l} f_{l,m} (\nu) Y_{l,m} (\cos \theta, \phi) \\
Y_{l,m} ( \cos \theta, \phi) & = & \sqrt{ \frac{2 l +1}{4 \pi} \frac{ (l-1)!}{(l+m)!}} P_{l}^m (\cos \theta) e^{i m \phi} \;.
\label{eq94}
\end{eqnarray}
Inserting the expanded distribution function into eq. \ref{fI} and for the "scattering in'' term we obtain:
\begin{eqnarray}
\displaystyle \frac{\partial f}{\partial \tau} (\nu_1, \hat{z}) \bigg|_{in} & = & \frac{3}{32 \pi^2} \sum_{l=0}^{\infty} \sum_{m=-l}^{l} \int d \beta_e d \mu_e d \phi_e \int d \mu_2 d \phi_0 \frac{n_{12} f_e (\beta_e)}{n_{22}^2 \gamma_e}  \nonumber \\ 
&& \times \bigg[ 1 + \eta_{12} \bigg( 1 + \frac{\eta_{12}}{2} \bigg) \bigg] f_{l,m} (\nu_2) \sqrt{\frac{2 l +1}{4 \pi}}  \nonumber \\
& & \times \frac{(l-m)!}{(l+m)!} P_{l}^m (\mu_2) e^{i m (\phi_0 + \phi_e)}  \nonumber \\
&=&\frac{3}{16 \pi} \sum_{l=0}^{\infty} \sqrt{\frac{2l +1}{4 \pi}} \int d \beta_e d \mu_e \int d \mu_2 d \chi_0 f_e (\beta_e)  \nonumber\\
&& \times \frac{ 2 n_{12}^2 n_{22}^2 + 2 n_{12} n_{22} ( \mu_2 + 1 ) + {(\mu_2 + 1 )}^2}{n_{12} n_{22}^4 \gamma_e \sqrt{1-\chi_0^2}} f_{l,0} (\nu_2) P_l^0 (\mu_2) \;. \nonumber \\
& &
\label{fani}
\end{eqnarray}
\newline
The integration over $\phi_e$ eliminates all the terms in $m \neq 0$. 
One can adopt an approach similar to the previous one by using the variables introduced in eq. \ref{wrights}, but we will use another set of variables introduced as follows:
\begin{eqnarray}
&&s=\ln \bigg(\frac{n_{12}}{n_{22}}\bigg) \nonumber\\
&&t=\ln \big( n_{12} n_{22}\big) \\
&&\mu_{0} = \frac{\mu_2 - 1}{n_{12} n_{22}} +1 \;.\nonumber
\label{eq96}
\end{eqnarray}
\newline
Substituting these variables into eq. \ref{fani}  and subtracting the ''scattering out'' term (see eq. \ref{fout}), we obtain a set of equations similar to eq. \ref{shift}
\begin{eqnarray}
\displaystyle \frac{\partial f}{\partial \tau} (\nu_1, \hat{z}) & = & \sum_{l=0}^{l=\infty} \int_{-\infty}^{\infty} P_{l,0} (s) f_{l,0} (e^s \nu_1) \ d s - f (\nu_1, \hat{z}) \nonumber \\
\displaystyle P_{l,0} (s) & = & \int_{\sinh {|s| \over 2}}^{1} f_e (p_e)\ P_{l,0} (s,p_e)\ d p_e \\
\displaystyle P_{l,0} (s,\beta_e) & = & -\frac{3}{64 \pi} \sqrt{\frac{2 l + 1}{\pi}} \frac{e^{{3 s} \over 2}}{\gamma_e^3 \beta_e^2} \int_{-t_0}^{t_0} e^{{t \over 2}} d t \int_{A-B}^{A+B} \frac{1+\mu_0^2}{\sqrt{ B^2 - {\big( A-\mu_0 \big)}^2}}  \nonumber \\
& & \times P_{l}^0 \big( e^t (\mu_0 -1) + 1\big)\ d \mu_0 \;, \nonumber
\label{eq97}
\end{eqnarray}
where
\begin{eqnarray}
&&t_0 = |s| - \ln \bigg(\frac{1+\beta_e}{1-\beta_e}\bigg) \nonumber \\
&&A = \frac{e^{-t}}{\gamma_e^2 \beta_e^2} \bigg[ 1 + \gamma_e^2 e^t - 2 \gamma_e e^{{t \over 2}} \cosh { \frac{s}{2}}\bigg] \label{eq98}\\
&&B = 2 \frac{e^{{t \over 2}}}{\gamma_e^2 \beta_e^2} \sqrt{ \bigg[ \cosh \bigg(\frac{s-t}{2}\bigg)-\gamma_e\bigg] \bigg[ \cosh \bigg(\frac{s+t}{2}\bigg)-\gamma_e\bigg]} \;. \nonumber
\end{eqnarray}
\newline
The function $P_{l,0} (s ,p_e)$ is just the function $P_{l,0} (s,\beta_e)$ with the $\beta_e$ substituted in terms of $p_e$ as well as $\gamma_e$. One can see here that to each value of $l$ one can associate a scattering kernel or a redistribution function $P_{l,0} ( s )$. The scattering kernel associated with the monopole term is actually related to the scattering kernel for the isotropic case as follows:
\begin{equation}
\displaystyle \displaystyle P_{0,0} (s)= \frac{1}{\sqrt{4\pi}} P_1 (s) \;.
\label{eq99}
\end{equation}
The scattering kernels for each $l$ value conserve the property written as follows:
\begin{equation}
P_{l,0} ( -s ) = e^{-3 s } P_{l,0} ( s ) \;.
\label{eq100}
\end{equation}
The change in the intensity for each value of $l$ can then be computed as follows:
\begin{eqnarray}
\frac{\partial I}{\partial \tau} ( x , \hat{z} ) & = & \sum_{l=0}^{\infty} \int_{-\infty}^{\infty} P_{l,0} ( s ) I_{l,0} ( e^{-s} x ) ds - I ( x , \hat{z} ) \nonumber \\
& = &  \sum_{l=0}^{\infty}\bigg[ \int_{-\infty}^{\infty} P_{l,0} ( s ) I_{l,0} ( e^{-s} x ) ds -\sqrt{\frac{2 l + 1 }{4 \pi }} I_{l,0} ( x )\bigg] \;,
\label{R}
\end{eqnarray}
where 
\begin{eqnarray}
I ( x , \theta, \phi ) & = & \sum_{l=0}^{\infty} \sum_{m=-l}^{l} I_{l,m} ( x ) Y_{l,m} ( \theta, \phi )  \nonumber \\
& = & 2 (k T_0)^3 \left[ \frac{x^3}{e^x - 1 } + \frac{e^x x^4}{(e^x - 1)^2} \sum_{l=2}^{\infty} \sum_{m=-l}^{l} a_{l,m} Y_{l,m} ( \theta, \phi ) \right] \;.
\label{eq102}
\end{eqnarray}
In eq. \ref{R} we have used the relation
\begin{equation}
P^0_l ( 1 ) = 1.
\label{eq104}
\end{equation}
We show in Fig. \ref{redistributionfunctionth} the scattering kernel $P_{l,m}$ for a thermal electron distribution for different temperatures and in Fig. \ref{redistributionfunctionspl} for a non-thermal electron distribution for different minimum momentum $p_1$ following a power law with index $\alpha=2.5$. 

\subsection{CMB multipoles and polarization of the SZ effect}

Now we proceed to derive the Stokes parameters $Q$ and $U$ for an anisotropic incident radiation. The Stokes parameter $Q$ can be written as follows:
\begin{eqnarray}
\frac{1}{\nu_1^3} \frac{\partial Q}{\partial \tau} ( \nu_1 ) & = & -\frac{3}{64 \pi^2} \sum_{l=0}^{\infty} \sum_{m=-l}^{l} \int d \beta_e d \mu_e d \phi_e \frac{f_e ( \beta_e )}{\gamma_e} \int \frac{ d\mu_2 d\phi_0}{n_{12} n_{22}^4}  \nonumber \\
&& \times \big[\cos (2 \phi_0 + 2 \phi_e ) \sin^2 (\theta_2 ) n_{12}^2 + 2 \cos ( \phi_0 + 2 \phi_e )  \nonumber \\
& & \times \sin (\theta_2) \sin (\theta_e) n_{12} r_{12} \gamma_e \beta_e + \cos (2\phi_e) \sin^2 (\theta_e) r_{12}^2 \beta_{e}^2 \gamma_e^2 \big]  \nonumber \\
&& \times \sqrt{\frac{2 l + 1}{4 \pi} \frac{(l-m)!}{(l+m)!}} e^{ i m ( \phi_0 + \phi_e )} P^m_l ( \mu_2 ) f_{l,m} \bigg( \frac{n_{12}}{n_{22}} \nu_1 \bigg) \;. 
\label{Qfirst}
\end{eqnarray}
Upon integration with respect to $ \phi_e$, only the terms with $m= \pm 2$ survive and we make use of the following property of the associated Legendre Polynomials 
\begin{equation}
P^{-m}_l (\mu ) = (-1)^m \frac{(l-m)!}{(l+m)!} P^m_l (\mu) \;,
\label{eq106}
\end{equation}
and we also impose the following condition on the photon redistribution function
\begin{equation}
f^*_{l,m} (\nu) = (-1)^m f_{l,-m} (\nu) \;.
\label{eq107}
\end{equation}
\begin{figure}
\begin{center}
\begin{tabular}{c c}
\includegraphics[width=80mm,height=75mm]{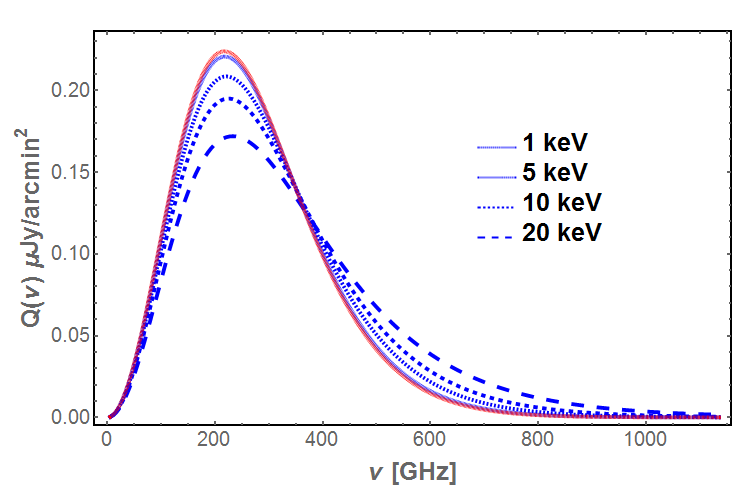} & \includegraphics[width=80mm,height=75mm]{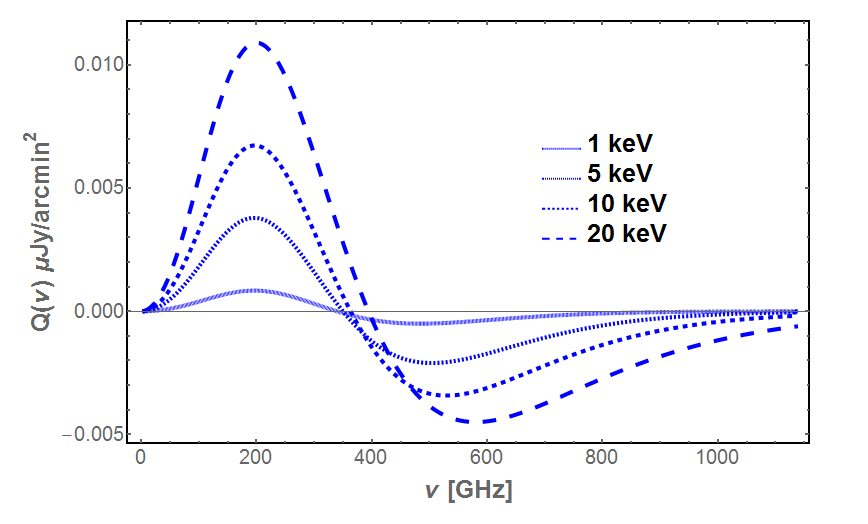}
\end{tabular}
\caption{The spectrum of the Stokes parameter $Q$ for different temperatures of a thermal electron distribution arising from the quadrupole (left) and octupole (right) of the CMB, assumed here to be $ a_{2,2} = 1.3 \times 10^{-5}$ and $ a_{3,2} = 8.7 \times 10^{-6}$, respectively. The red curve represents the non-relativistic $Q$. The optical depth value is $\tau=0.001$.}
\label{Q22Q32th}
\end{center}
\end{figure}
\begin{figure}
\begin{center}
\begin{tabular}{c c}
\includegraphics[width=80mm,height=75mm]{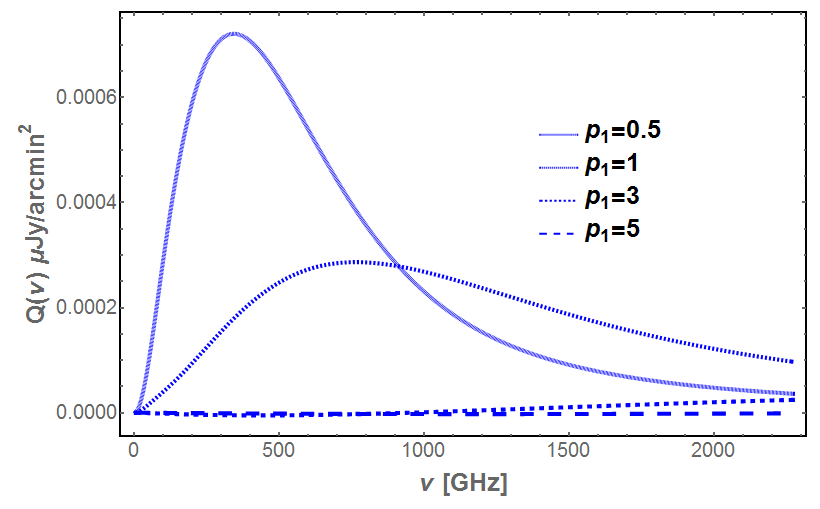} & \includegraphics[width=80mm,height=75mm]{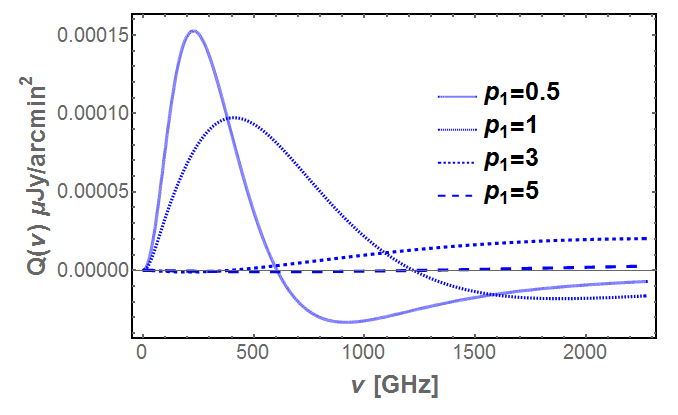}
\end{tabular}
\caption{The spectrum of the Stokes parameter $Q$ for the quadrupole (left) and octupole (right) in the case of a single power law distribution of electrons of spectral index $\alpha=2.5$. The quadrupole of the CMB is assumed here to be $ a_{2,2} = 1.3 \times 10^{-5}$ and that of the octupole to be $ a_{3,2} = 8.7 \times 10^{-6}$. The optical depth value is $\tau=1\times10^{-5}$.}
\label{Q22Q32spl}
\end{center}
\end{figure}
\begin{figure}
\begin{center}
\begin{tabular}{c c}
\includegraphics[width=80mm,height=75mm]{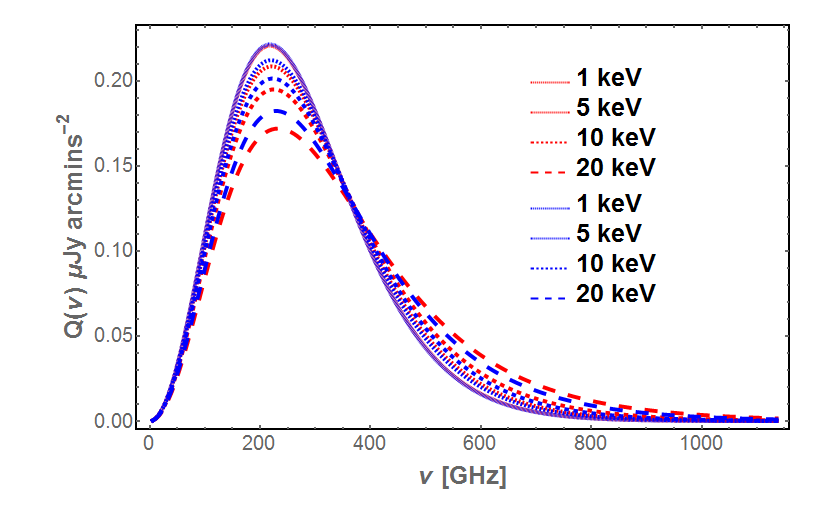} & \includegraphics[width=80mm,height=75mm]{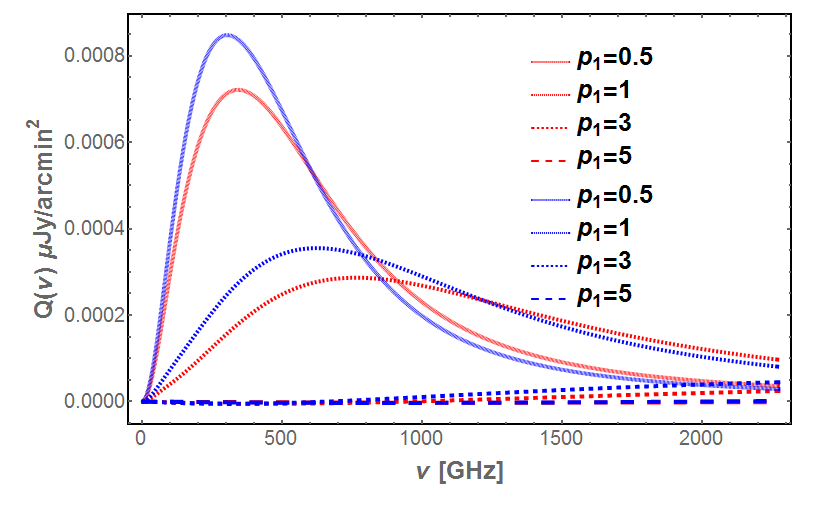}
\end{tabular}
\caption{The spectrum of the Stokes parameter $Q$ for the superposition of the CMB quadrupole and octupole (blue curve) for a thermal electron distribution (left panel) and for a non-thermal electron distribution (right panel). Optical depth values of $10^{-3}$ and $10^{-5}$ have been used  for the thermal and non-thermal distributions, respectively. The spectral index of the power-law distribution is  $\alpha=2.5$. The red curves represent the spectrum of $Q$ where the contribution from the quadrupole only is considered.}
\label{Q2232overall}
\end{center}
\end{figure}
The Stokes parameter $Q$ then is written as follows:
\begin{eqnarray}
\frac{1}{\nu_1^3} \frac{\partial Q}{\partial \tau} ( \nu_1 ) & = & -\frac{3}{16 \pi} \sum_{l=2}^{\infty} \sqrt{\frac{2 l +1}{4 \pi}\frac{(l-2)!}{(l+2)!}} \int d \beta_e d \mu_e \frac{f_e (\beta_e)}{\gamma_e}  \nonumber \\
& & \times \int \frac{d \mu_2 d \chi_0}{n_{12} n_{22}^4 \sqrt{1-\chi_0^2}} \big[(1-\mu_2^2) n_{12}^2 + 2 n_{12} \beta_e \gamma_e (\mu_2 - 1)  \nonumber \\
& & \times \chi_0 \sqrt{(1-\mu_2^2) (1-\mu_e^2)} + \beta_e^2 \gamma_e^2 (\mu_2-1)^2 (1-\mu_e^2) (2 \chi_0^2 - 1 )\big]  \nonumber \\
& & \times Re \big[ f_{l,2} \bigg(\frac{n_{12}}{n_{22}} \nu_1 \bigg) \big] P^2_l ( \mu_2 ) \;.
\label{eq108}
\end{eqnarray}
Similarly the Stokes parameter $U$ can be written like the previous one
\begin{eqnarray}
\frac{1}{\nu_1^3} \frac{\partial U}{\partial \tau} ( \nu_1 ) &=& \frac{3}{16 \pi} \sum_{l=2}^{\infty} \sqrt{\frac{2 l +1}{4 \pi}\frac{(l-2)!}{(l+2)!}} \int d \beta_e d \mu_e \frac{f_e (\beta_e)}{\gamma_e}\int \frac{d \mu_2 d \chi_0}{n_{12} n_{22}^4 \sqrt{1-\chi_0^2}}  \nonumber \\
& & \times \big[(1-\mu_2^2) n_{12}^2 + 2 n_{12} \beta_e \gamma_e (\mu_2 - 1) \chi_0 \sqrt{(1-\mu_2^2) (1-\mu_e^2)}  \nonumber \\
& & + \beta_e^2 \gamma_e^2 (\mu_2-1)^2 (1-\mu_e^2) (2 \chi_0^2 - 1 )\big] \nonumber \\
& & \times Im \big[ f_{l,2} \bigg(\frac{n_{12}}{n_{22}} \nu_1 \bigg) \big] P^2_l ( \mu_2 ) \;.
\label{eq109}
\end{eqnarray}
These expressions can actually be simplified further into the following equations similar to those used to compute the intensity $I$
\begin{eqnarray}
\frac{1}{\nu_1^3} \frac{\partial Q}{\partial \tau} ( \nu_1 ) & = & \sum_{l=0}^{\infty} \int_{-\infty}^{\infty} P_{l,2} ( s ) Re \big[ f_{l,2} ( e^s \nu_1 ) \big] d s \nonumber \\
\frac{1}{\nu_1^3} \frac{\partial U}{\partial \tau} ( \nu_1 ) & = & -\sum_{l=0}^{\infty} \int_{-\infty}^{\infty} P_{l,2} ( s ) Im \big[ f_{l,2} ( e^s \nu_1 ) \big] d s  \;,
\end{eqnarray}
where
\begin{eqnarray}
P_{l,2} ( s ) & = & \int_{\sinh (|s|/2)}^ {\infty} P_{l,2} ( s , p_e ) f_e ( p_e ) d p_e \nonumber \\
&& \nonumber \\
P_{l,2} ( s,\beta_e ) & = & -\frac{3}{32 \pi} \sqrt{\frac{2 l + 1}{4 \pi} \frac{(l-2)!}{(l+2)!}} \frac{e^{\frac{3}{2} s }}{\gamma_e^2 \beta_e^2} \int_{-t_0}^{t_0} e^{\frac{t}{2}} d t  \nonumber \\
&& \nonumber \\
&& \times \int_{A-B}^{A+B} d \mu_0 \frac{P^2_l ( 1 + e^t ( \mu_0 -1) )}{\sqrt{B^2 -(A-\mu_0)^2}} \frac{ \mu_0 - 1 }{2 + e^t (\mu_0 - 1 )}  \nonumber \\
&& \nonumber \\
&& \times \bigg[(\mu_0-1) \big[ 2 - e^t \big( \gamma_e^2 (\mu_0 - 1) ( 1+\beta_e^2)-2 \big) \big]  \nonumber \\
& & - 8 \gamma_e (\mu_0 - 1) e^{t/2} \cosh (\frac{s}{2}) -4 \cosh s \bigg] \;.
\label{TheQUspectrum}
\end{eqnarray}
The variables $ A$,$ B$ and $t$ are given in eq.\ref{eq98}. The redistribution kernel $P_{l,2} (s)$ follows a similar kind of relationship as that of $P_{l,0} (s)$ written as follows
\begin{equation}
P_{l,2} (-s) = e^{-3 s } P_{l,2} ( s ) \;.
\label{eq111}
\end{equation}
This allows us to cast the Stokes parameters $Q$ and $U$ in terms of the incident radiation intensity and for completeness we also include the intensity Stokes parameter $I$
\begin{eqnarray}
&&\frac{\partial I}{\partial \tau} ( x , \hat{z} )=  \sum_{l=0}^{\infty}\bigg[ \int_{-\infty}^{\infty} P_{l,0} ( s ) I_{l,0} ( e^{-s} x ) ds -\sqrt{\frac{2 l + 1 }{4 \pi }} I_{l,0} ( x )\bigg],\nonumber \\
&&\frac{\partial Q}{\partial \tau} (x) = \sum_{l=2}^{\infty} \int_{-\infty}^{\infty} P_{l,2} ( s ) Re \big[ I_{l,2} ( e^{-s} x ) \big] d s, \nonumber \\
&&\frac{\partial U}{\partial \tau} (x) = -\sum_{l=2}^{\infty} \int_{-\infty}^{\infty} P_{l,2} ( s ) Im \big[ I_{l,2} ( e^{-s} x ) \big] d s \;.
\label{stokesprimor}
\end{eqnarray}
\begin{figure}
\begin{center}
\begin{tabular}{c c}
\includegraphics[width=80mm,height=75mm]{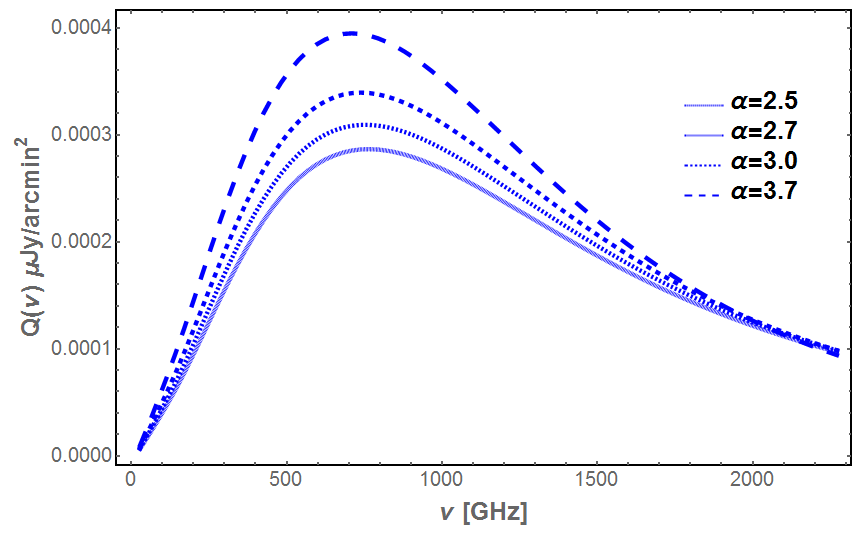} & \includegraphics[width=80mm,height=75mm]{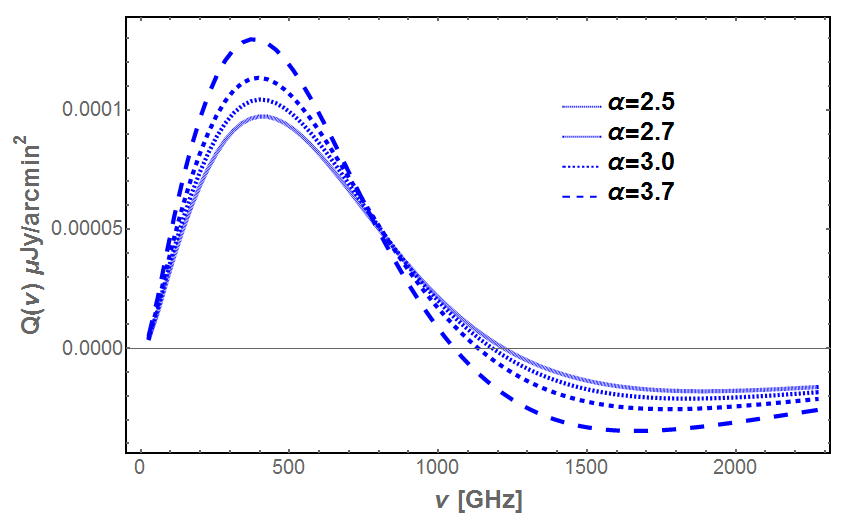}
\end{tabular}
\caption{The spectrum of the Stokes parameter $Q$ for the CMB quadrupole and octupole computed for different spectral index $\alpha$ of a single power-law distribution of electrons. The values of the minimum momentum and optical depth here are $p_1=1$ and $1\times 10^{-5}$, respectively.}
\label{spectra2232}
\end{center}
\end{figure}
\begin{figure}
\begin{center}
\begin{tabular}{c c}
\includegraphics[width=80mm,height=75mm]{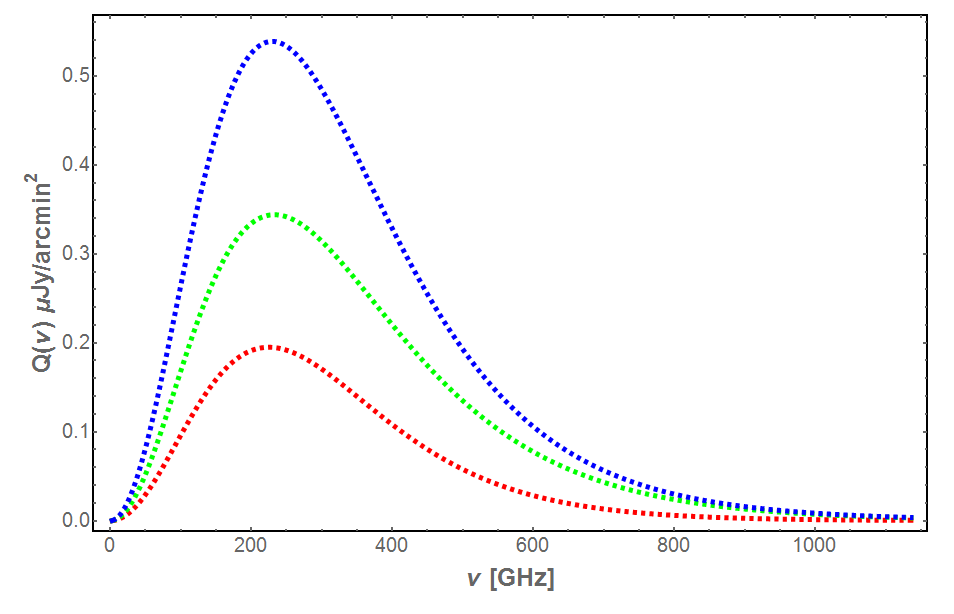} & \includegraphics[width=80mm,height=75mm]{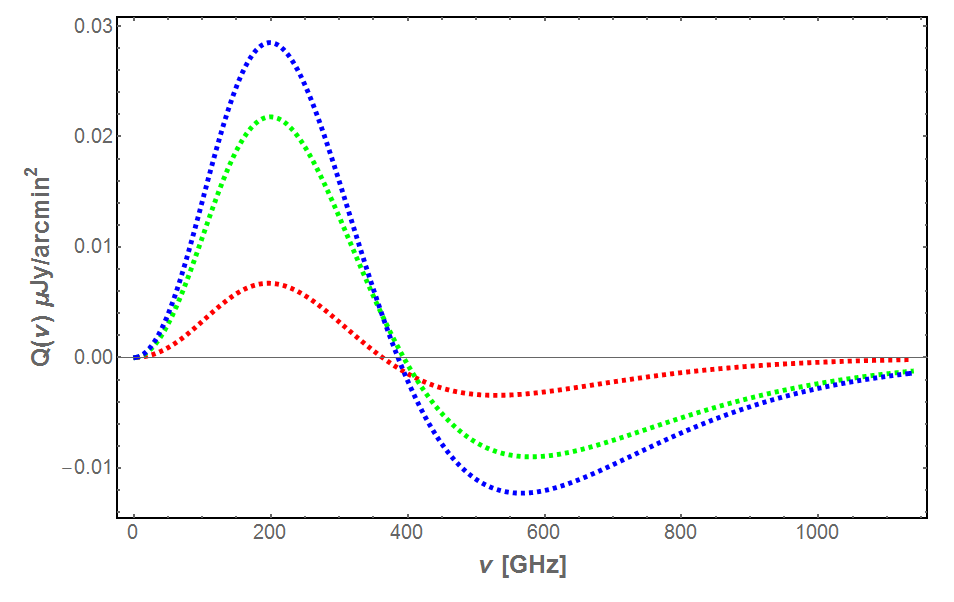}
\end{tabular}
\caption{The spectrum of the Stokes parameter $Q$ for the quadrupole (left panel) and the octupole (right panel) computed for a combination of two thermal electron populations (blue curves): a thermal electron population with $kT=10$ keV and $\tau=0.001$ (red curves) and another thermal electron population with $kT=20$ keV and $\tau=0.002$ (green curves).}
\label{thermalthermal}
\end{center}
\end{figure}
\begin{figure}
\begin{center}
\begin{tabular}{c c}
\includegraphics[width=80mm,height=75mm]{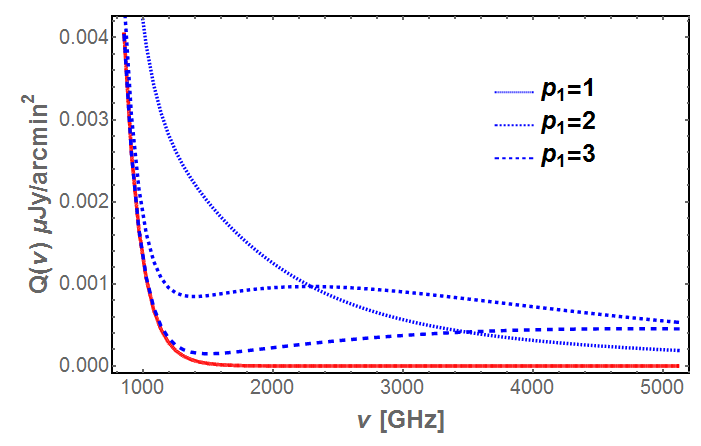} & \includegraphics[width=80mm,height=75mm]{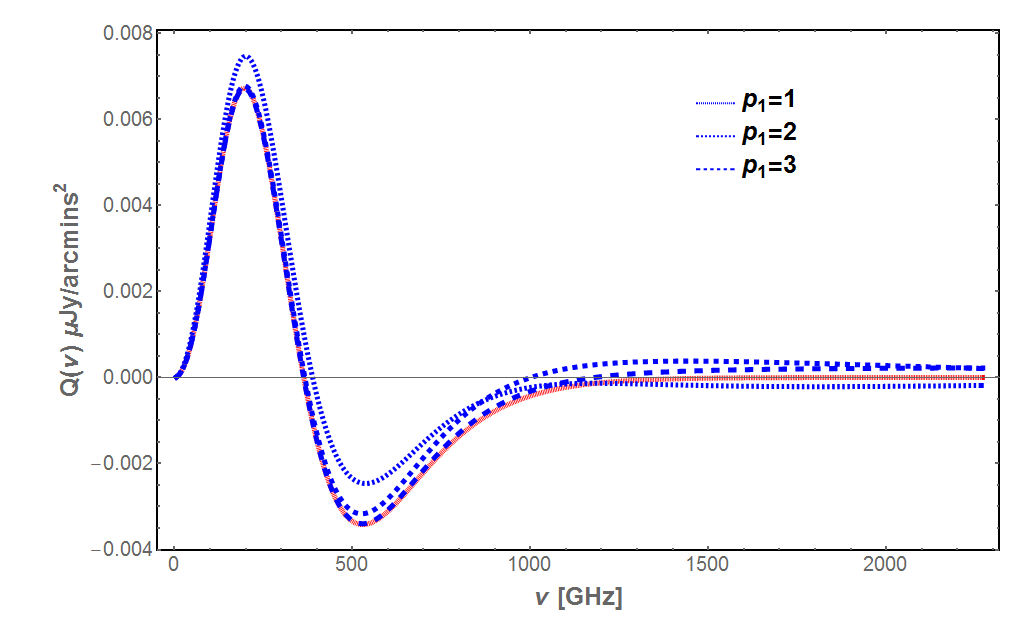}
\end{tabular}
\caption{The spectrum of the Stokes parameter Q for the CMB quadrupole (left panel) and octupole (right panel) for a thermal electron distribution with $kT=10$ keV and $\tau=10^{-3}$ (red curves) combined with a power-law electron distribution, spectral index $\alpha=2.5$ and $\tau=10^{-4}$, for different minimum momentum $p_1$. The blue curves represent the resulting spectrum for different values of $p_1$.}
\label{thermalspl}
\end{center}
\end{figure}

In addition to that, one can also extend eq. \ref{stokesprimor} to compute the spectrum for a combination of electron populations. We show the derivation in the Appendix and we just present the result here,
\begin{eqnarray}
\displaystyle \frac{\partial I}{\partial \tau} ( x , \hat{z} ) & = & \frac{\tau_A}{\tau} \sum_{l=0}^{\infty} \int_{-\infty}^{\infty} P_{l,0,A} ( s ) I_{l,0} ( e^{-s} x ) ds + \nonumber \\
& & +  \frac{\tau_B}{\tau} \sum_{l=0}^{\infty} \int_{-\infty}^{\infty} P_{l,0,B} ( s ) I_{l,0} ( e^{-s} x ) ds  -\sqrt{\frac{2 l + 1 }{4 \pi }} I ( x , \hat{z} ), \nonumber \\
\displaystyle \frac{\partial Q}{\partial \tau} ( x , \hat{z} ) & = & \frac{\tau_A}{\tau} \sum_{l=0}^{\infty} \int_{-\infty}^{\infty} P_{l,2,A} ( s ) Re [I_{l,2} ( e^{-s} x )] ds + \nonumber \\
& & +  \frac{\tau_B}{\tau} \sum_{l=0}^{\infty} \int_{-\infty}^{\infty} P_{l,2,B} ( s ) Re [I_{l,2} ( e^{-s} x )] ds \nonumber \\
\displaystyle \frac{\partial U}{\partial \tau} ( x , \hat{z} ) & = & -\frac{\tau_A}{\tau} \sum_{l=0}^{\infty} \int_{-\infty}^{\infty} P_{l,2,A} ( s ) Im [I_{l,2} ( e^{-s} x )] ds + \nonumber \\
& & -  \frac{\tau_B}{\tau} \sum_{l=0}^{\infty} \int_{-\infty}^{\infty} P_{l,2,B} ( s ) Im [I_{l,2} ( e^{-s} x )] ds \nonumber,\\
\label{ABQU}
\end{eqnarray}
where $\tau_A$ and $\tau_B$ are the optical depths of electron population A and B, respectively.

\section{Polarization spectra}

Eq. \ref{stokesprimor} in the previous Sect. 3.5 allow us to compute the Stokes parameters $Q$ and $U$ for any value of $l$. In this work we computed $Q$ and $U$ only up to $l=3$. We show in Fig. \ref{Q22Q32th} the spectrum of the Stokes  parameter $Q$ arising from the quadrupole and the octupole of the CMB for different temperatures of a thermal plasma residing in galaxy clusters. The first thing that one notices is that in the relativistic case, higher multipoles of the CMB contribute to polarization contrary to the non-relativistic case where only the CMB quadrupole contributes. The CMB quadrupole and the octupole show also distinct and different spectral features. Relativistic effects become more pronounced for hot clusters while for low electron temperatures, $kT <1$ keV, the relativistic spectrum and the non-relativistic one become nearly identical in the case of the CMB quadrupole.

The peak of the $Q$ spectrum for the CMB quadrupole still peaks roughly around same frequency, $\approx 216$ GHz, but with a slight deviation towards higher frequencies, reaching 230 GHz for 20 keV,  and the peak intensity value is lowered with increasing electron temperature. Between a cluster at 20 keV and a cluster at 1 keV, the difference is $\approx 0.05$ $\mu$Jy/arcmin$^2$ at 216 GHz for an optical depth of $\tau=0.001$.

\begin{figure}
\begin{tabular}{c}
\includegraphics[width=80mm,height=60mm]{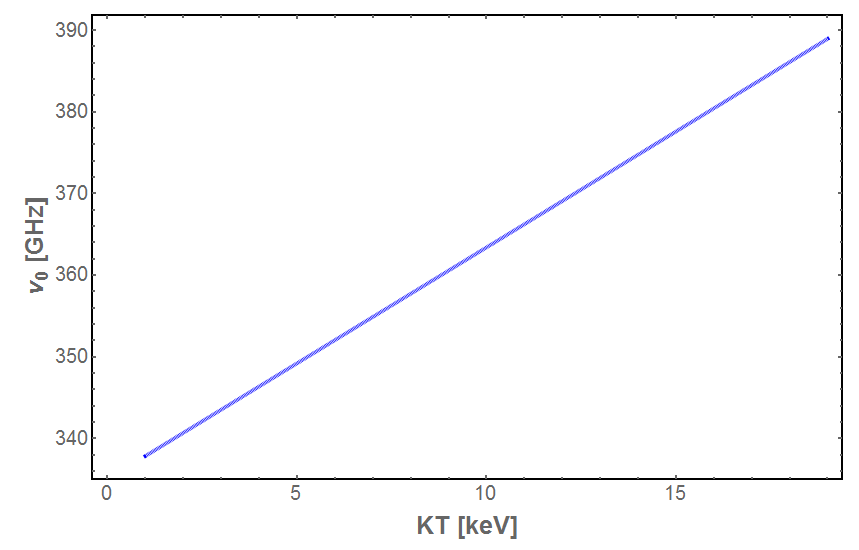}\\
\includegraphics[width=80mm,height=60mm]{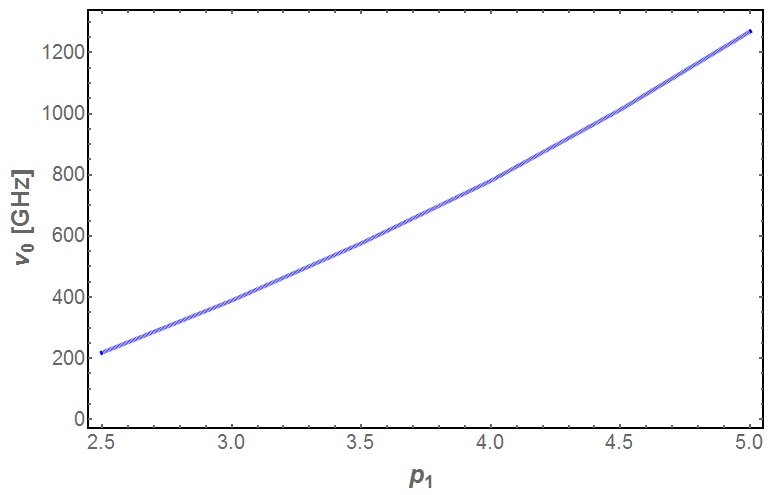}\\
\includegraphics[width=80mm,height=60mm]{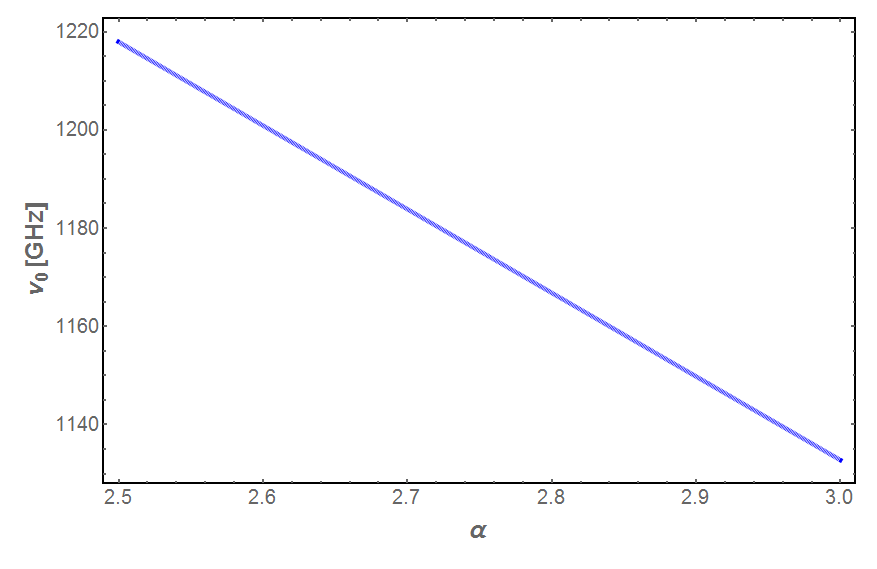}
\end{tabular}
\caption{The variation of the cross-over frequency $\nu_0$ as a function of the electron temperature  (top panel) for the case of a thermal distribution of electrons, of the minimum normalized momentum $p_1$ for a non-thermal electron distribution (mid panel) for a fixed $\alpha=2.5$, and of the spectral index $\alpha$ (bottom panel) in the case of a power-law electron distribution for a fixed $p_1=1$.}
\label{nu0}
\end{figure}

In the case of the CMB octupole, the amplitude of the spectrum is smaller and increases with the electron temperature. The $Q$ spectrum for the CMB octupole shows a cross-over frequency which takes the value $\approx 340$ GHz in the thermal case for $kT=1$ keV and it increases with electron temperature; for a temperature of 20 keV it is found at $\approx 396$ GHz. This cross-over frequency means a change in the polarization state, e.g. changing it from vertically polarized to horizontally polarized.
We show in Fig \ref{nu0} (top panel) the variation of the cross-over frequency $\nu_0$ as a function of the temperature of a thermal gas. The cross-over frequency varies with the temperature as $\nu_0=[335+2.84 (kT/keV)]$ GHz.
This relationship can be used, in principle, to measure the electron plasma temperature provided that the crossover frequency is measured with sufficient frequency accuracy.

We note that the CMB octupole term contributes to the total polarization spectrum like a perturbation of the main contribution that is given by the CMB quadrupole term. To illustrate this fact, we show in Fig. \ref{Q2232overall} (left panel) the spectrum of the total Stokes parameter $Q$ from the superposition of the CMB quadrupole and octupole for a thermal electron distribution. The presence of the CMB octupole-induced contribution makes the peak of the total $Q$ spectrum higher at the frequency of the maximum of the CMB quadrupole-induced spectrum, reaching values of  0.182 $\mu$Jy instead of 0.172 $\mu$Jy for $kT=20$ keV. At frequency higher than $\approx 371$ GHz, the total $Q$ spectrum becomes lower in amplitude w.r.t. the case of the CMB quadrupole-induced contribution due to the negative amplitude of the CMB octupole-induced term in this frequency range.
Therefore, the contribution of the CMB octupole term is small but not negligible, and one could consider to use the cross-over frequency of the CMB octupole term to measure the cluster electron temperature.

To separate the contribution from the CMB quadrupole and octupole, one can use 
the cluster (electron plasma) temperature derived from X-ray spectral observations to fit the
CMB octupole-induced SZ effect polarization and therefore disentangle its contribution from the CMB quadrupole-induced SZ effect polarization in the total $Q$ Stoke parameter spectrum: the prior knowledge of the cluster temperature allows to measure the cross-over frequency $\nu_0$ and then estimate at that frequency the intrinsic CMB quadrupole-induced contribution to the SZ effect polarization which coincides with the total $Q$.
Once the CMB quadrupole-induced term is derived, it can be subtracted from the total $Q$ spectrum thus disentangling the CMB octupole-induced term.  

In addition to the thermal $Q$ spectrum, we compute also the spectrum of the polarized SZ effect for non-thermal electrons which has been done here for the first time. We show in Fig \ref{Q22Q32spl} the spectrum of the Stokes parameter $Q$ for different values of the minimum momentum $p_1$ and for a single power-law electron distribution with spectral index $\alpha=2.5$ which is representative of the observed spectra in radio-halos and radio-galaxies. In the non-thermal case, the amplitude of the spectrum is lower than in the thermal case because of the lower optical depth of this non-thermal plasma and the amplitude changes with the minimum momentum $p_1$. The spectrum in this case extends however to much higher frequencies depending on the values of $p_1$.

The CMB octupole-induced term to $Q$ has been also derived here for the first time and, similarly to the thermal case, it shows the presence of a cross-over frequency determining a change in polarization.
We show in Fig \ref{nu0} the variation of the cross over frequency with respect to $p_1$ (mid panel) for $p_1 >2.5$. Interestingly, two values of the cross-over frequency are seen for $p_1$ greater than $\approx 2$: in fact, for $p_1=3$, the cross-over frequencies are found at 389 GHz and 4000 GHz. We find that the relationship between the cross-over frequency $\nu_0$ and $p_1$ is not linear but quadratic in $p_1$ and it can be described by a polynomium $\nu_0=(-284.3+ 93.4 p_1+43.4 p_1^2)$ GHz in the range $p_1 \approx 2.5-5.0$. As in the thermal plasma case, this relation can be used to derive the value of $p_1$ given a sufficient frequency resolution of the CMB octupole-induced spectrum. We also show in Fig. \ref{Q2232overall} (right panel) the spectrum of the total non-thermal Stokes parameter $Q$ from the superposition of the CMB quadrupole and the octupole contributions to the total $Q$.

Furthermore, in Fig \ref{spectra2232} we show how the spectrum changes with different electron spectral index $\alpha$. We find that softer spectral index leads to higher amplitude of the $Q$ spectrum. For a given value of the momentum $p_1$,  the cross-over frequency of the CMB octupole-induced spectrum depends on the spectral index $\alpha$, with a softer spectral index leading to lower values of the cross-over frequency. We show in Fig \ref{nu0} (bottom panel) the relationship between the cross-over frequency $\nu_0$ and the spectral index $\alpha$. A linear relationship is found which can be reproduced by the relation $\nu_0 =(1644-170.3 \alpha)$ GHz. This can be used to measure $\alpha$ if the value of $p_1$ is known.  The combination of the dependence of the cross-over frequency on $p_1$ and $\alpha$  can be used to set constraints on these spectral parameters of the non-thermal electron distribution.

Using eq. \ref{ABQU}, we also compute the spectra for different combinations of electron populations. In Fig. \ref{thermalthermal}, we show the resulting spectrum (blue curve) for a combination of two thermal electron distributions. The overall spectrum is the superposition of the individual spectra. Another scenario which is interesting to look at is the combination of a thermal electron population with  a non-thermal one. We show in Fig. \ref{thermalspl} the resulting spectrum (blue curve) at high frequencies for a thermal electron distribution combined with a non-thermal electron population, for different values of the miminum momentum $p_1$. We obtain that in this case, because of the low number density of non-thermal electrons, the non-thermal spectrum does not affect the thermal spectrum at low frequencies except for low values of the momentum $p_1 (<1)$. Therefore at low frequencies, the resulting spectrum is dominated by the thermal spectrum. However, at high frequencies ($>$1000 GHz), the non-thermal spectrum dominates and therefore the resulting spectrum is entirely non-thermal. Interestingly, the existence of a non-thermal distribution superposed to a thermal distribution has an impact on the value of the cross-over frequency, $\nu_0$ of the CMB octupole-induced spectrum. Without a non-thermal distribution of electrons, we obtain $\nu_0=361$ GHz for $kT=10$ keV, while in the presence of a non-thermal electron distribution with $p_1=1$, $\nu_0=389$ GHz, $\nu_0=366$ GHz  with $p_1=2$, and  $\nu_0=361$ GHz with $p_1=3$. The presence of a non-thermal distribution of electrons with $p_1>1$ produces an additional cross-over frequency $\nu_0$, e.g. at $\approx$ 1000 GHz for $p_1=2$ and $\approx$ 1191 GHz for $p_1=3$. It is also noticed that the cross-over frequency depends on the minimum momentum $p_1$ of the non-thermal distribution, lower values of $p_1$ lead to higher cross-over frequencies. As in the case of a thermal electron distributions, it is possible to measure $p_1$ from the CMB octupole-induced spectrum at high-frequencies and then disentangle the two electron populations in a galaxy cluster involving non-thermal activities.

\subsection{Application to the Bullet cluster}

In order to assess the detectability of the SZ effect polarization arising from the multipoles of the CMB radiation, we compute the spectrum of the Stokes parameter $Q$ for the case of the Bullet cluster. SZ effect measurements in intensity have been done for this cluster over a wide frequency range: with ACBAR at 150 and 275 GHz \cite{Gomez2004}, with the SEST telescope at 150 GHz \cite{Andreani1999}, with APEX at 150 GHz \cite{Halverson2009}, with the SPT at 150 GHz \cite{Plagge2010}, with ATCA \cite{Malu2010} at 18 GHz and with Herschel-SPIRE at 600, 850 and 1200 GHz \cite{Zemcov2010}. The availability of multifrequency data allowed the determination of multiple components arising in the SZ effect signal \cite{CM2011, CM2015}. These last authors found that the signal of the main sub-cluster is better explained using two electron components, a thermal electron distribution having optical depth $\tau=1.1 \times 10^{-2}$ and $kT= 14.2$ keV that is co-spatial with a non-thermal electron distribution with optical depth $3\times 10^{-4}$, $p_1=1$ and spectral index $\alpha=3.7$ \cite{CM2015}. In the following, we compute the SZ effect polarization using these parameters and we calculate the flux integrated over a region of radius 5 arcmin from the center of this cluster.
\begin{figure}
\includegraphics[width=150mm,height=120mm]{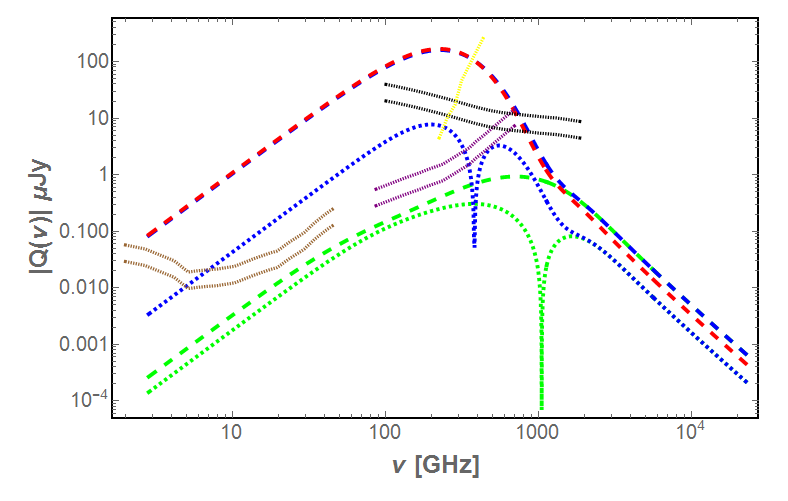}
\caption{The polarization spectrum (dashed-red) of the Bullet cluster calculated over 5 arcmin$^2$. The dashed- blue curve is the quadrupole spectrum whereas the dotted-blue is that of the octupole. The brown, purple and the black curves represent the sensitivy of SKA, ALMA and Millimetron for 260 and 1000 hrs of integration. The yellow curve represents the sensitivity of CORE++.}
\label{fullspectrum2232}
\end{figure}

We show in Fig .\ref{fullspectrum2232}, the polarization spectrum (red curve) up to $10^4$ GHz computed for the Bullet cluster  from the superposition of the CMB quadrupole (dashed-blue) and the octupole (dotted-blue) signals. The polarization signal (red) reaches a maximum of 160 $\mu$Jy at 215 GHz. At 383 GHz, the signal is not contaminated by the CMB octupole contribution as this is the cross-over frequency of the CMB octupole-induced spectrum, and for frequencies $>$ 383 GHz, the signal goes below that of the CMB quadrupole-induced one. This shows how the CMB octupole contributes constructively with the CMB quadrupole before the cross-over frequency ($\approx$ 383 GHz) and destructively for frequencies above that. Using eq. \ref{degPol} we  estimate the degree of polarization at some specific frequencies and we obtained for 10 GHz $\Pi=1.65 \times 10^{-6}$, for 200 GHz $\Pi=3 \times 10^{-6}$ and for 1000 GHz $\Pi=3 \times 7.7 \times 10^{-4}$.

In order to highlight the impact of the non-thermal component, we show in Fig. \ref{fullspectrum2232} the spectrum of the CMB quadrupole-induced  (dashed-green) and the CMB octupole-induced  (dotted-green) terms arising from the non-thermal plasma component. The low value of $p_1$ in this particular case of the Bullet cluster makes the non-thermal component not completely negligible over the entire spectrum. At frequencies $<2000$ GHz, the thermal component dominates whereas at frequencies $>$ 2000 GHz, the spectrum is entirely non-thermal. We also show in Fig. \ref{fullspectrum2232} the sensitivities of SKA (brown)  \cite{Carilli2008, Dewney2013}, ALMA (purple) \cite{Carilli2008} and Millimetron (black) \cite{millimetron} for 260 and 1000 hrs of integration. The sensitivities of SKA and ALMA are at 1$\sigma$ while that of Millimetron is at 5$\sigma$. It is important to stress that these polarization sensitivities are estimated by assuming that the Stokes $Q$ sensitivity is a factor of $\sqrt{2}$ higher than the Stokes $I$ sensitivity. We have also shown the CORE++ sensitivity (yellow) as reported in \cite{CORE++}.

\section{Discussion and conclusions}

In this paper we derived the polarization of the SZ effect in a relativistic approach by solving the relativistic polarized Boltzmann equation in the case of the CMB photons and of electrons that reside in galaxy clusters and radio galaxy lobes. Under the Thomson limit assumption, the Stokes parameters $Q$ and $U$ have been derived. We found that when relativistic effects are taken into account, all the CMB multipoles are implicated in the production of polarization, at variance with the non-relativistic case where only the CMB quadrupole is involved. Contrary to previous works on the SZ effect \cite{CL1998}, we have calculated explicitly the spectrum of the Stokes parameter induced by the CMB octupole. The spectrum of this last term shows distinct spectral features compared to the one induced by the CMB quadrupole, in particular the existence of a cross-over frequency $\nu_0$. Contribution from multipoles higher than the CMB octupole are expected in our general approach but they haven't been considered in our work since their amplitudes would be quite small and the resulting signal would be well below the detection limit of any current and future instruments. 

We also found that the spectral feature of the SZ effect polarization spectrum depends on the electron distribution causing the IC scattering. The spectral feature of the SZ effect polarization is completely different between thermal and non-thermal electrons. In the case of thermal electrons, the spectral features of the $Q$ spectrum (quadrupole/octupole) depend on the electron temperature whereas in the case of the non-thermal electrons with a power-law distribution, the spectral features depend on $p_1$ and $\alpha$. In the particular case of the CMB octupole-induced term, the cross-over frequency depends on the temperature (e.g. 340 GHz for 1 keV, 365 GHz for 10 keV, 396 GHz for 20 keV) for a thermal electron distribution and on $p_1$ as well as $\alpha$ for that of a power-law electron distribution. We have briefly discussed the source of biases in the determination of the cluster temperature and of the electron minimum momentum from the measurements of the cross-over frequency $\nu_0$, as well as the possibility to disentangle the CMB quadrupole and octupole terms from a prior measurement of the cluster temperature (or the minimum momentum of a non-thermal electron distribution).

In addition to that, we have computed the polarization spectrum arising from a combination of electron populations which has not been done before in previous works \cite{CL1998}. The resulting spectrum is the superposition of individual spectra, e.g. in the case of two thermal electron distribution that exist co-spatially, the polarization spectrum is amplified by the distribution with higher temperature. In the particular case of the CMB octupole-induced  spectrum, the cross-over frequency is moved towards higher frequency. For the case of a thermal electron distribution co-spatially existing with a non-thermal electron distribution, the latter only causes an impact on the polarization spectrum of both the CMB quadrupole and octupole-induced terms at higher frequencies. However for low values of $p_1$ ($<1$), the non-thermal electron distribution impact can also be seen at lower frequencies. Again in the case of the CMB octupole-induced term, the cross-over frequency is shifted depending on the value of the momentum $p_1$ and the spectral index $\alpha$. 

The ability to compute the SZ effect polarization using two electron distributions allowed us to compute the total SZ effect polarization spectrum of the Bullet cluster using the parameters derived from multifrequency SZ effect observations in intensity.
The spectrum of the Bullet cluster shows interesting frequency ranges over which polarization arising from the CMB multipoles can be explored. In the frequency range 30 GHz to 700 GHz, the SZ effect polarization spectrum can be measured with a sensitivity to the Stokes parameter $Q$  of $\approx$ 10 $\mu$Jy. This frequency range is suitable since it can be considered clean from the foreground polarized synchrotron emission. This falls into the frequency coverage of ALMA and Millimetron, which cover approximately the ranges 86 GHz up to 750 GHz and 100 GHz to 1800 GHz, respectively. Observing at $\approx$ 88 GHz where the sensitivity of ALMA is maximum, i.e. $\approx$ 0.3 $\mu$Jy, would render the SZ effect polarization signal measurable. The detection limit of ALMA goes below both the predicted CMB quadrupole and the octupole-induced signals, which would render both signals observable. The distinct spectral features of the CMB quadrupole and the octupole-induced spectrum would also allow them to be disentangled using multifrequency observations. For clusters with electron temperature measurements available through X-ray or SZ effect intensity measurements, a strategy that can be employed in achieving this is by observing at the cross-over frequency of the CMB octupole-induced term, e.g. around 390 GHs for Bullet cluster. At this frequency the polarization spectrum consist of only the CMB quadrupole-induced term without any contribution from the CMB octupole-induced term (being zero at its cross-over frequency). Then in order to recover the latter, one can measure the SZ effect polarization spectrum at another frequency where the CMB quadrupole-induced term  spectrum can be subtracted to retain the CMB octupole-induced one. However high spectral resolution would be required in order to observe the signal with sufficient precision around the cross-over frequency. 

Promising experimental possibilities to measure the SZ polarization are offered by future experiments: the SKA can reach a sensitivity of 0.01 $\mu$Jy at 5 GHz and down to 0.1 $\mu$Jy at 45 GHz for 1000 hrs of integration time. This would allow SZ effect polarization spectrum to be measured with high accuracy at low frequencies. 
Unfortunately low frequency measurements can suffer from the polarized synchrotron emissions coming from the radio halos or other sources. Nevertheless the distinctive spectral feature of the latter would allow it to be removed through multifrequency measurements \cite{Hall2014}. In addition, if an r.m.s. value of the CMB quadrupole or the CMB octupole is what we are interested in, then the polarized synchrotron emissions would cancel upon averaging over several clusters \cite{Hall2014}.This is so because the synchrotron polarization angle will not correlate from cluster to cluster. Hence combining ALMA and the SKA, the CMB quadrupole and the octupole-induced terms can be determined. The averaging process will also reduce the kinematic SZ effect polarization assuming the peculiar velocities of cosmic structures are uncorrelated \cite{Hall2014}.

The SZ effect polarization coming from non-thermal electrons reveals that polarization can also be searched in the extended lobes of radio galaxies where IC emission have been observed. However high sensitivities ($\approx$ 0.01 $\mu$Jy at 20 GHz, $\approx$ 0.33 $\mu$Jy at 243 GHz ) would be needed to at least measure the CMB quadrupole-induced spectrum. This would require at least 5000 hrs of integration time for SKA and ALMA. 

Another nuisance for both low and high frequency observations of SZ effect polarization from cosmic structures would be the background E-mode polarization of the CMB itself. In order to estimate the contribution of the E-mode to the SZ effect polarization, we assume that the CMB quadrupole/octupole is same at every location in the universe. For angular scale of typical cluster, $\approx$ 5 arcmin that corresponds to $\approx l \approx2000$, we compare the value of the amplitude of the E-mode to the CMB temperature quadrupole/octupole. We found in this way that the CMB E-mode contribute only to $\approx$ 0.3 \% to the SZ effect polarization.

We finally discuss in the following the comparison between the results obtained in our work and previous ones \cite{CL1998, Itoh1998}. 
The first difference between our work and previous ones is that we used a general formalism which allowed us to compute the SZ effect polarization for any electron distribution, whereas in previous works a series expansion in terms of a thermal electron distribution temperature has been used to study the thermal SZ effect polarization only.

We stress also that we have been able to compute the SZ effect polarization spectra for both thermal and non-thermal electron distributions in a complete relativistic approach without relying on any approximation (apart from the Thomson scattering approximation which is justified in this study). In particular, the non-thermal SZ effect polarization has been presented here for the first time because of its relevance in search of SZ effect polarization in the radio haloes of galaxy clusters and lobes of radio-galaxies.

Furthermore we have also been able to compute the SZ effect polarization for a general combination of electron populations with different nature, which is a relevant case for galaxy clusters containing non-thermal plasma either in the form of radio halos/relics and extended jets of radio galaxies.

We also stress here that we have computed for the first time the spectrum of the CMB octupole-induced polarization which was not considered  in previous works on SZ effect polarization. We have shown that the CMB octupole-induced contribution is not negligible in the case of high temperature clusters as well as in lobes of radio-galaxies where relativistic electrons are present. We have shown also that the spectrum of the CMB octupole-induced polarization shows an interesting spectral feature, i.e. a cross-over frequency, which can be used to estimate the electron temperature (for a thermal electron population), or the minimum momentum $p_1$ and the spectral index $\alpha$ of non-thermal electron distributions. We also discussed how multifrequency observations, by taking advantage of the CMB octupole's cross-over frequency, would allow one to disentangle the CMB quadrupole and octupole-induced spectrum.

Finally, we have also computed the complete polarization spectrum expected from the Bullet cluster using parameters derived from multifrequency observations of the SZ effect intensity. In the context of expectations from observed clusters, we have shown that telescopes like the SKA, ALMA,  Millimetron and CORE++ have the sensitivity to measure the polarization spectrum from a typical Bullet-like cluster. A statistical study of the SZ effect polarization signals from a sample of high-T clusters will be presented elsewhere and will point to cosmological applications of this technique in large-scale observations of the polarized cosmic microwave background.

\section*{Acknowledgments}

This work is based on the research supported by the South African Research Chairs Initiative of the Department of Science and Technology and National Research Foundation of South Africa (Grant No 77948).
M.S.E. and P.M. acknowledge support from the DST/NRF SKA post-graduate bursary initiative.\\
\textit{Disclaimer: any opinion, finding and conclusion or recommendation expressed in this material is that of the author(s) and the NRF does not accept any liability in this regard}.

\appendix
\section{Appendix. SZ polarization spectrum for combination of electron populations}

Using the formalism presented in this paper one can also compute the SZ effect polarization due to two electron populations residing in the same ICM. 
This was done by \cite{CM2003} for the case of intensity only while here we will do it for the Stokes parameter 
$Q$ and $U$ in the case of the quadrupole and octopole. We write the distribution function of the electron $f_e ( p )$ as follows:
\begin{equation}
f_e ( p ) = C_A f_{e,A} ( p ) + C_B f_{e,B} ( p ),
\label{eq114}
\end{equation}
where $f_{e,A} ( p )$ corresponds to the distribution function of electron population A and $f_{e,B} ( p )$ corresponds to 
the distribution function of electron population B. $C_A$ and $C_B$ are normalization constants with $C_A + C_B = 1$ \cite{CM2003} and 
\begin{eqnarray}
&\displaystyle C_A = \frac{\tau_A}{\tau} \nonumber \\
& \displaystyle C_A = \frac{\tau_B}{\tau},
\label{eq115}
\end{eqnarray}
with $\tau = \tau_A + \tau_B$. The total scattering kernel for any value of $m$ and $l$ due to the combination of population A and B is written as follows:
\begin{eqnarray}
\displaystyle P_{l,m} ( s ) &=& \int f_{e} ( p )\ P_{l,m} ( s , p)\ d p \nonumber \\
\displaystyle &=& \int  C_A f_{e,A} ( p )  P_{l,m} ( s , p) + C_B f_{e,B} ( p )  P_{l,m} ( s , p)\ d p \nonumber \\
 \displaystyle &=& C_A P_{l,m, A} ( s ) + C_B P_{l,m,B} ( s ) .
\label{eq116}
\end{eqnarray}
The Stokes parameters $I$ can be written as:
\begin{eqnarray}
\displaystyle \frac{\partial I}{\partial \tau} ( x , \hat{z} ) & = & \frac{\tau_A}{\tau} \sum_{l=0}^{\infty} \int_{-\infty}^{\infty} P_{l,0,A} ( s ) I_{l,0} ( e^{-s} x ) ds + \nonumber \\
& & +  \frac{\tau_B}{\tau} \sum_{l=0}^{\infty} \int_{-\infty}^{\infty} P_{l,0,B} ( s ) I_{l,0} ( e^{-s} x ) ds  -\sqrt{\frac{2 l + 1 }{4 \pi }} I ( x , \hat{z} ), \nonumber \\
& & \label{appendixABI}
\end{eqnarray}
and the Stokes parameters $Q$ and $U$ are written as:
\begin{eqnarray}
\displaystyle \frac{\partial Q}{\partial \tau} ( x , \hat{z} ) & = & \frac{\tau_A}{\tau} \sum_{l=0}^{\infty} \int_{-\infty}^{\infty} P_{l,2,A} ( s ) Re [I_{l,2} ( e^{-s} x )] ds + \nonumber \\
& & +  \frac{\tau_B}{\tau} \sum_{l=0}^{\infty} \int_{-\infty}^{\infty} P_{l,2,B} ( s ) Re [I_{l,2} ( e^{-s} x )] ds \nonumber \\
\displaystyle \frac{\partial U}{\partial \tau} ( x , \hat{z} ) & = & -\frac{\tau_A}{\tau} \sum_{l=0}^{\infty} \int_{-\infty}^{\infty} P_{l,2,A} ( s ) Im [I_{l,2} ( e^{-s} x )] ds + \nonumber \\
& & -  \frac{\tau_B}{\tau} \sum_{l=0}^{\infty} \int_{-\infty}^{\infty} P_{l,2,B} ( s ) Im [I_{l,2} ( e^{-s} x )] ds \nonumber .\\
\label{appendixABQU}
\end{eqnarray}
The resulting spectrum of the Stokes parameters for a combination of electron populations is the superposition of their individual spectrum.

\end{document}